%% file: HIN-15-006_temp.tex
\pdfoutput=1

\documentclass[11pt,twoside,a4paper,cmspaper,final,collab]{cms-tdr}
\def\svnVersion{391124}\def\cmsCernNoTag{CERN-EP/2016-105}\def\cmsCernDate{\today}\def\cmsMessage{Published in Physics Letters B as \href{http://dx.doi.org/10.1016/j.physletb.2017.01.075}{\doi{10.1016/j.physletb.2017.01.075.}}}

\begin{document}\cmsNoteHeader{HIN-15-006}

\hyphenation{had-ron-i-za-tion}
\hyphenation{cal-or-i-me-ter}
\hyphenation{de-vices}
\RCS$Revision: 390864 $
\RCS$HeadURL: svn+ssh://svn.cern.ch/reps/tdr2/papers/HIN-15-006/trunk/HIN-15-006.tex $
\RCS$Id: HIN-15-006.tex 390864 2017-03-02 22:17:31Z ztu $
\newlength\cmsFigWidth
\ifthenelse{\boolean{cms@external}}{\setlength\cmsFigWidth{\columnwidth}}{\setlength\cmsFigWidth{0.8\textwidth}}
\ifthenelse{\boolean{cms@external}}{\providecommand{\cmsLeft}{top\xspace}}{\providecommand{\cmsLeft}{left\xspace}}
\ifthenelse{\boolean{cms@external}}{\providecommand{\cmsRight}{bottom\xspace}}{\providecommand{\cmsRight}{right\xspace}}
\newcommand {\roots}{\ensuremath{\sqrt{s}}\xspace}
\newcommand {\rootsNN}{\ensuremath{\sqrt{s_{_{NN}}}}\xspace}
\newcommand{\noff}{\ensuremath{N_\text{trk}^\text{offline}}\xspace}
\providecommand{\EPOS}{\textsc{epos}\xspace}
\newcommand{\pp}{\ensuremath{\Pp\Pp}\xspace}
\newcommand{\PbPb}{\text{PbPb}\xspace}
\newcommand{\AuAu}{\text{AuAu}\xspace}
\newcommand {\pPb}{\ensuremath{\Pp\text{Pb}}\xspace}
\newcommand{\KET}{\ensuremath{\mathrm{KE}^\mathrm{T}}\xspace}
\newcommand{\KETavg}{\ensuremath{\langle\mathrm{KE}_\mathrm{T}}\rangle\xspace}

\cmsNoteHeader{HIN-15-006}
\title{Multiplicity and rapidity dependence of strange hadron production in pp, pPb, and PbPb collisions at the LHC}

\date{\today}

\abstract{
Measurements of strange hadron (\PKzS, \PgL+\PagL, and \PgXm+\PagXp) transverse momentum spectra
in \pp, \pPb, and \PbPb collisions are presented over a wide range of rapidity and event charged-particle multiplicity. The data were collected with the CMS detector at the CERN LHC
in \pp  collisions at $\roots = 7$\TeV, \pPb collisions at $\rootsNN = 5.02\TeV$, and \PbPb collisions at $\rootsNN = 2.76\TeV$.
The average transverse kinetic energy is found to increase with multiplicity, at
a faster rate for heavier strange particle species in all systems. At similar multiplicities, the difference in average transverse kinetic energy
between different particle species is observed to be larger for \pp  and \pPb events than for \PbPb events.
In \pPb collisions, the average transverse kinetic energy is found to be slightly larger in the
Pb-going direction than in the p-going direction for events with large multiplicity. The spectra are compared to models motivated by hydrodynamics.
}

\hypersetup{%
pdfauthor={CMS Collaboration},%
pdftitle={Multiplicity and rapidity dependence of strange hadron production in pp, pPb, and PbPb collisions at the LHC},%
pdfsubject={CMS},%
pdfkeywords={CMS, physics, heavy ion, spectra, radial flow}}

\maketitle

\section{Introduction}
\label{sec:intro}

Studies of strange-particle production
in high energy collisions of protons and heavy ions provide important
means to investigate the dynamics of the collision process.
Earlier studies of relativistic heavy ion collisions
at the BNL RHIC and CERN SPS colliders indicated an enhancement of strangeness production with respect to proton-proton (\pp) collisions~\cite{Andersen:1998vu,STAR}, which was historically interpreted to be due to the formation of a
high-density quark-gluon medium~\cite{Rafelski:1982pu}. The abundance of strange particles at different center-of-mass energies is in line with calculations from thermal statistical models~\cite{Abelev:2007xp, Andersen:1999ym, Adams:2003fy}. In gold-gold (\AuAu) collisions at RHIC, strong azimuthal correlations of final-state hadrons were observed, suggesting that the produced medium behaves like a near-perfect fluid undergoing a pressure-driven anisotropic expansion~\cite{STAR}. Studies of strangeness and light flavor production and dynamics in heavy ion collisions have provided further insight into the medium's fluid-like nature and evidence for its partonic collectivity~\cite{STAR,PHENIX}.

In recent years, the observation of a long-range ``ridge" at small azimuthal separations in two-particle correlations in \pp~\cite{Khachatryan:2010gv}
and proton-lead (\pPb)~\cite{Khachatryan:2012dih,atlas:2012fa,alice:2012qe}
collisions with high event-by-event charged-particle multiplicity (referred to hereafter as ``multiplicity") has provided an indication for collective effects in systems that are an order of magnitude smaller in size than heavy ion collisions. The nature of the observed
long-range particle correlations in high multiplicity \pp  and \pPb collisions
is still under intense debate~\cite{Dusling:2015gta}. While the collective flow of a fluid-like medium
provides a natural interpretation~\cite{Bozek:2011if,Bozek:2012gr,Bzdak:2013zma,Werner:2010ss}, other models attribute this behavior to the initial correlation of gluons~\cite{Dusling:2012wy,Dusling:2012cg,Dumitru:2014yza,Gyulassy:2014cfa,Li:2012hc}, or the anisotropic escape of particles~\cite{He:2015hfa}.

Studies of identified particle production and correlations in high multiplicity \pp  and \pPb collisions provide detailed information about the underlying particle
production mechanism. Identified particle (including strange-hadron)
transverse momentum (\pt) spectra and azimuthal anisotropies in lead-lead (\PbPb) collisions at the CERN LHC have been studied~\cite{Abelev:2013vea,Abelev:2014pua} and
described by hydrodynamic models~\cite{Song:2013qma, Zhu:2015dfa}.
Similar measurements have been performed in \pPb collisions
as a function of multiplicity, where an indication of a common velocity
boost to the produced particles, known as ``radial flow''~\cite{Abelev:2013haa,Chatrchyan:2013eya},
and for a mass dependence of the anisotropic flow~\cite{ABELEV:2013wsa,Khachatryan:2014jra} have been observed.
When comparing \pPb and \PbPb systems at similar multiplicities, a stronger
radial velocity boost is seen in the smaller \pPb collision system~\cite{Khachatryan:2014jra, Abelev:2013haa}.
This could be related to a much higher initial energy density in a high multiplicity
but smaller system, resulting in a larger pressure gradient outward along the radial direction,
as predicted in Ref.~\cite{Shuryak:2013ke}. To perform a quantitative comparison, a common
average radial-flow velocity from different collision systems can be extracted from
a simultaneous fit to the spectra of various particle species,
based on the blast-wave model~\cite{Schnedermann_blastwave}. Inspired by hydrodynamics, the blast-wave model assumes a common kinetic freeze-out temperature and radial-flow velocity for all particles during the expansion of the system. The dependence of spectral shapes for identified hadrons on the multiplicity has been observed in high energy electron and
proton-antiproton collisions~\cite{Albajar:1989an, Alexopoulos:1988na}, but this observation was not explored extensively in the hydrodynamic context. The blast-wave fit has been studied in \pp, deuterium-gold, and \AuAu collisions at RHIC~\cite{Abelev:2008ab}. In \pp\ collisions, it has been shown through studies with simulation that color reconnection processes could describe the observed multiplicity dependence of identified particle spectra~\cite{Ortiz:2013yxa, Abelev:2013vea}.

It is of interest to study possible collective phenomena in very high multiplicity \pp  collisions,
as demonstrated by the observation of long-range particle correlations in these events~\cite{Khachatryan:2010gv}.
Since \pp  events represent an even smaller system than \pPb events, a stronger radial-flow boost
might be present compared to \pPb and \PbPb events at a comparable multiplicity~\cite{Shuryak:2013ke}.
Furthermore, in a \pPb collision, the system is not symmetric in pseudorapidity ($\eta$).
If a fluid-like medium is formed, its energy density could be different on the p- and Pb-going
sides, which could lead to an asymmetry in the collective radial-flow effect as a function of $\eta$.
Hydrodynamical models predict that the average \pt (or, equivalently, the average transverse kinetic energy $\KETavg$, where $\KETavg \equiv \langle m_\mathrm{T}\rangle - m$, with $m_\mathrm{T} = \smash[b]{\sqrt{m^{2} + \pt^{2}}}$ and $m$ the particle mass) of produced particles is larger in the
Pb-going direction than in the p-going direction, while this trend could be reversed
in models based on gluon saturation~\cite{Bozek:2013sda}.
Measurement of identified particle \pt spectra as a function of $\eta$ could thus
help to constrain theoretical models.

This Letter presents measurements of strange-particle \pt spectra in \pp, \pPb, and \PbPb collisions as a function of the multiplicity in the events. Specifically, we examine the spectra of \PKzS, \PgL, and \PgXm\ particles, where the inclusion of the charge-conjugate states is implied for \PgL\ and \PgXm\ particles. The data were collected with the CMS detector at the LHC. With the implementation of a dedicated high-multiplicity trigger,
the \pp  and \pPb data samples exhibit multiplicities comparable to that
observed in peripheral \PbPb collisions, where ``peripheral'' refers to ${\sim}50$--100\% centrality, with centrality defined as the fraction of the total inelastic cross section. The most central collisions have 0\% centrality. This overlap in mean multiplicity allows the three systems, with drastically different collision geometries, to be compared. The large solid-angle coverage of the CMS detector permits the strange-particle \pt spectra to be studied in different rapidity ranges, and thus the study
of possible asymmetries with respect to the p- and Pb-going directions in \pPb collisions.

\section{Detector and data samples}
\label{sec:detector}

The central feature of the CMS apparatus is a superconducting solenoid of 6\unit{m} internal diameter, which provides an axial field of 3.8 T. Within the solenoid volume are a silicon pixel and strip tracker (with 13 and 14 layers in the central and endcap regions, respectively),
a lead tungstate crystal
electromagnetic calorimeter (ECAL), and a brass and scintillator hadron
calorimeter (HCAL), each composed of a barrel and two endcap sections. The tracker covers the pseudorapidity range $\abs{\eta}<2.5$. Reconstructed tracks with $1<\pt<10\GeV$ typically have resolutions of 1.5--3\% in \pt and 25--90 (45--150)\micron in the transverse (longitudinal) impact parameter~\cite{TRK-11-001}. The ECAL and HCAL each cover $\abs{\eta}<3.0$ while forward hadron calorimeters (HF) cover $3<\abs{\eta}<5$.  Muons with $\abs{\eta}<2.4$ are measured with gas-ionization detectors embedded in the steel flux-return yoke outside the solenoid.
A more detailed description of the CMS detector, together with a definition
of the coordinate system and the relevant kinematic variables,
can be found in Ref.~\cite{Chatrchyan:2008zzk}.
The Monte Carlo (MC) simulation
of the particle propagation and detector response is based on the \GEANTfour~\cite{GEANT4} program.

The data samples used in this analysis are as follows:
\pp  collisions collected in 2010 at $\roots = 7\TeV$, \pPb collisions collected in 2013 at $\roots = 5.02\TeV$, and \PbPb collisions collected in 2011 at $\rootsNN = 2.76\TeV$, with integrated luminosities of 6.2\pbinv, 35\nbinv, and 2.3\mubinv, respectively.

For the \pPb data, the beam energies are 4\TeV for the protons and 1.58\TeV per nucleon for the lead nuclei. The data were collected in two different run periods: one with the protons circulating in the clockwise direction in the LHC ring, and one with them circulating in the counterclockwise direction. By convention, the proton beam rapidity is taken to be positive when combining the data from the two run periods.
Because of the asymmetric beam conditions, the nucleon-nucleon center-of-mass in the \pPb collisions moves with speed $\beta = 0.434$ in the laboratory frame, corresponding to a rapidity of 0.465.
As a consequence, the rapidity of a particle in the nucleon-nucleon
center-of-mass frame ($y_\text{cm}$) is detected in the laboratory frame ($y_\text{lab}$) with
a shift, $y_\text{lab} = y_\text{cm} + 0.465$.
The \pPb particle yields reported in this Letter are presented in terms of $y_\text{cm}$, rather than $y_\text{lab}$, for better correspondence with the results from the \pp  and \PbPb collisions.
\section{Selection of events and tracks}

The triggers, event reconstruction, and event selection are the same as those discussed for \pp, \pPb, and \PbPb collisions in Refs.~\cite{Chatrchyan:2013nka,Khachatryan:2010gv}.
They are briefly outlined in the following paragraphs for \pp  and \pPb collisions,
which are the main focus of this Letter. A subset of peripheral \PbPb data collected in 2011 with a minimum-bias trigger is reprocessed using the same event selection and track
reconstruction algorithm as for the present \pPb and \pp  analyses,
in order to more directly compare the three systems at the same multiplicity.
Details of the 2011 \PbPb analysis can be found in Refs.~\cite{Chatrchyan:2012xq,Chatrchyan:2013nka}.

Minimum-bias \pPb events are triggered by requiring at least one track
with $\pt > 0.4$\GeV to be found in the pixel tracker.
Because of hardware limitations in the data acquisition rate, only a small fraction
$({\sim}10^{-3})$ of triggered minimum-bias events are recorded.
In order to collect a large sample of high-multiplicity \pPb collisions, a dedicated
high-multiplicity trigger is implemented using the CMS Level-1 (L1) and high-level
trigger (HLT) systems~\cite{Adam:2005zf}. At L1, the total transverse energy summed over the ECAL and HCAL is required to exceed either 20 or 40\GeV, depending on the multiplicity requirement as specified below. Charged particles are reconstructed at the HLT level using the pixel detectors. It is required that these tracks originate within a cylindrical region (30\unit{cm} in length along the direction of the beam axis and 0.2\unit{cm} in radius in the direction perpendicular to that axis) centered on the nominal interaction point. For each event, the number of pixel tracks (${N}_\text{trk}^\text{online}$) with
$\abs{\eta}<2.4$ and $\pt > 0.4\GeV$ is determined for each reconstructed vertex.
Only tracks with a distance of closest approach 0.4\unit{cm} or less to
one of the vertices are included. The HLT selection requires ${N}_\text{trk}^\text{online}$ for the vertex with the largest number of tracks to exceed a specific value. Data are collected
in \pPb collisions with thresholds ${N}_\text{trk}^\text{online}>100$ and 130 for events with an L1 transverse energy threshold of 20\GeV, and ${N}_\text{trk}^\text{online}>160$ and 190 for events with an L1 threshold of 40\GeV.
While all events with $N^\text{online}_\text{trk}>190$ are
accepted, only a fraction of the events from the other thresholds are
retained.  This fraction is dependent on the instantaneous luminosity. Data from both the minimum-bias
trigger and the high-multiplicity trigger are retained for offline
analysis. Similar high-multiplicity triggers, with different thresholds, were developed for \pp  collisions, with details given in Ref.~\cite{Khachatryan:2010gv}.

In the subsequent analysis of all collision systems, hadronic events are selected by requiring the presence of
at least one energy deposit larger than 3\GeV in each of the two HF calorimeters.
Events are also required to contain a primary vertex
within 15\unit{cm} of the nominal interaction point along the beam axis and 0.15\unit{cm} in the transverse direction, where the primary vertex is the reconstructed vertex with the largest track multiplicity. At least two reconstructed
tracks are required to be associated with this primary vertex, a condition that
is important only for minimum-bias events. Beam-related background is suppressed
by rejecting events in which less than 25\% of all reconstructed tracks satisfy
the high-purity selection defined in Ref.~\cite{TRK-11-001}. In the \pPb data sample, there is a 3\% probability to have at least one additional interaction in the same bunch crossing (pileup). The procedure used to reject pileup events in \pPb collisions is described in Ref.~\cite{Chatrchyan:2013nka}. It is based on the number of tracks associated with each reconstructed vertex and the distance between different vertices.  A purity of 99.8\% for single pPb collision events is achieved for the highest multiplicity \pPb range studied in this Letter. For the pp data, the average number of collisions per bunch crossing is 1.2. However, \pp  interactions that are well separated from each other do not interfere. Thus, among events identified as containing pileup, the event is retained if the separation between the primary vertex and any other vertex exceeds 1\unit{cm}. In such events, only tracks from the highest multiplicity vertex are used.

With the above criteria, 97\% (98\%) of the simulated \pPb events generated with the \EPOS\ \textsc{lhc}~\cite{Pierog:2013ria} (\HIJING\ 2.1~\cite{Gyulassy:1994ew}) programs are selected. Similarly, 94\% (96\%) of the \pp  events simulated with the \PYTHIA 6 Tune Z2~\cite{Sjostrand:2006za} (\PYTHIA 8 Tune 4C ~\cite{Sjostrand:2007gs}) programs are selected.

The event-by-event charged-particle multiplicity $N_\text{trk}^\text{offline}$ is defined using primary tracks, i.e, tracks that satisfy the high-purity criteria of Ref.~\cite{TRK-11-001} and, in addition, the following criteria designed to improve track quality and ensure the tracks emanate from the primary vertex. The impact parameter significance of the track with respect to the primary vertex in the direction along the beam axis, $d_z/\sigma(d_z)$, is required to be less than 3, as is the corresponding impact parameter in the transverse plane, $d_\mathrm{T}/\sigma(d_\mathrm{T})$. The relative \pt uncertainty, $\sigma(\pt)/\pt$, must be less than 10\%.
To ensure high tracking efficiency and to reduce the rate of misreconstructed
tracks, the tracks are required to satisfy $\abs{\eta}<2.4$ and $\pt > 0.4\GeV$. Based on simulated samples generated with the \HIJING\ program, the efficiency for primary track reconstruction is found to be greater than 80\% for charged particles with $\pt>0.6\GeV$ and $\abs{\eta}<2.4$. For the multiplicity range studied in this Letter, no dependence of the tracking efficiency on multiplicity is found and the rate of misreconstructed tracks is 1--2\%.

The \pp, \pPb, and \PbPb data are divided into classes based on $N_\text{trk}^\text{offline}$.  The quantity $N_\text{trk}^\text{corrected}$ is the corresponding multiplicity corrected for detector and algorithm inefficiencies in the same kinematic region ($\abs{\eta}<2.4$ and $\pt > 0.4\GeV$). The fraction of the total multiplicity found in each interval and the average number of tracks both before and after accounting for the corrections are listed in Table~\ref{tab:newmultbinning} for the \pp  data and in Ref.~\cite{Chatrchyan:2013nka} for the \pPb and \PbPb data. The uncertainty in the average value $\langle N_\text{trk}^\text{corrected} \rangle$ is evaluated from the uncertainty in the tracking efficiency, which is 3.9\% for a single track~\cite{CMS-PAS-TRK-10-002}. For the \pp  data, six multiplicity intervals, indicated in Table~\ref{tab:newmultbinning}, are defined, which are inclusive for the lower bounds and exclusive for the upper bounds, as indicated in Table~\ref{tab:newmultbinning}. The average $N_\text{trk}^\text{offline}$ value of minimum-bias events is similar to that for the multiplicity range $N_\text{trk}^\text{offline} < 35$. For the \pPb and \PbPb data, eight intervals are defined. These eight intervals are indicated, e.g., in the legend of Fig.~\ref{fig:V0spectra_midRapidity}. Note that, unlike \pp  and \PbPb collisions, \noff\ for
\pPb collisions is not determined in the center-of-mass frame.
However, the difference in the \noff\ definition between the laboratory and the center-of-mass
frames is found to be minimal and so this difference is ignored. The detector condition has
been checked to be stable for events with different multiplicities.

\begin{table*}[ht]\renewcommand{\arraystretch}{1.2}\addtolength{\tabcolsep}{-1pt}
\centering
\topcaption{ \label{tab:newmultbinning} Fraction of the full event sample
in each multiplicity interval and the average multiplicity per interval for \pp  data.
The multiplicities $N_\text{trk}^\text{offline}$ and $N_\text{trk}^\text{corrected}$ are determined for $\abs{\eta}<2.4$ and $\pt>0.4\GeV$ before and after efficiency corrections, respectively.
The third and fourth columns list the average values of $N_\text{trk}^\text{offline}$ and $N_\text{trk}^\text{corrected}$.}
\begin{tabular}{l*{4}{c}}
\hline
Multiplicity interval (\noff) & Fraction  & $\left<N_\text{trk}^\text{offline}\right>$ & $\left<N_\text{trk}^\text{corrected}\right>$ \\
\hline
$[0, 35)$ &0.93 & 12 & 14$\pm$1 \\
$[35, 60)$ &0.06 & 43 & 50$\pm$2 \\
$[60, 90)$ &$6 \times 10^{-3}$ & 68 & 79$\pm$3 \\
$[90, 110)$  &$2 \times 10^{-4}$ & 97 & 112$\pm$4 \\
$[110, 130)$ &$1 \times 10^{-5}$ & 116 & 135$\pm$5  \\
$[130, \infty)$ &$7 \times 10^{-7}$ & 137 & 158$\pm$6 \\
\hline
\end{tabular}
\end{table*}
\section{\texorpdfstring{The \PKzS, \PgL, and \PgXm}{K0S, Lambda/anti-Lambda, Cascade/anti-Cascade} reconstruction and yields}
\label{sec:V0}

The reconstruction and selection procedures for \PKzS, \PgL, and \PgXm\ candidates are presented in Refs.~\cite{Khachatryan:2011tm,Khachatryan:2014jra}.
To increase the efficiency for tracks with low momenta and large impact parameters,
both characteristic of the strange-particle decay products, the loose
selection of tracks, as defined in Ref.~\cite{TRK-11-001}, is used.
The \PKzS\ and \PgL\ candidates (generically referred to as ``$\mathrm{V}^{0}$s") are
reconstructed, by combining oppositely charged particles to define a secondary vertex. Each of the two tracks must have hits in at least four layers of the silicon tracker, and transverse and longitudinal impact parameter significances with respect to the primary vertex greater than 1. The distance of closest approach of the pair of tracks to each other is required to be less than
0.5\unit{cm}. The fitted three-dimensional vertex of the pair of tracks is required to
have a $\chi^{2}$ value divided by the number of degrees of freedom less than 7.
Each of the two tracks is assumed to be a pion in the case of the \PKzS\ reconstruction. As the proton carries nearly all of the momentum in the \PgL\ decay, the higher-momentum track is assumed to be a proton and the other track a pion in the case of the \PgL\ reconstruction. To reconstruct \PgXm\ particles, a \PgL\ candidate is combined with
an additional charged particle carrying the correct sign, to define a common secondary vertex. This additional track is required to have hits in at least four layers of the silicon tracker, and both the transverse and longitudinal impact parameter significances with respect to the primary vertex
are required to exceed 3.

\begin{figure*}[bht]
\centering
\includegraphics[width=\linewidth]{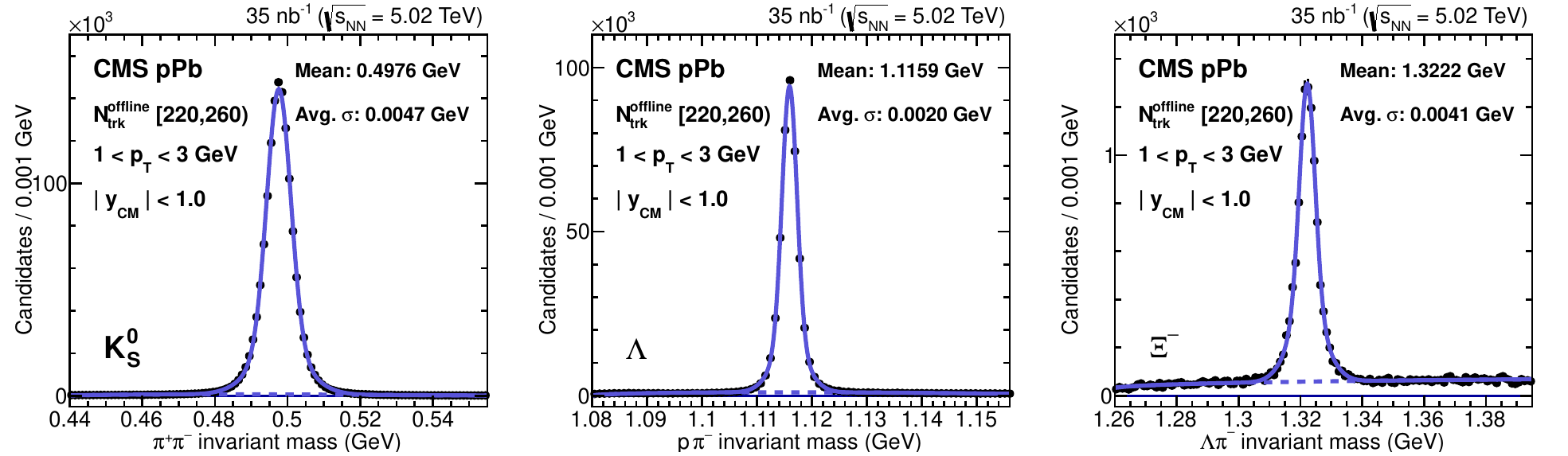}
  \caption{ \label{fig:v0mass} Invariant mass distribution of \PKzS\ (left),  \PgL\ (middle), and  \PgXm\ (right) candidates in the \pt range 1--3\GeV for $220 \leq \noff < 260$ in \pPb collisions. The inclusion of the charge-conjugate states is implied for \PgL\ and \PgXm\ particles. The solid lines show the results of fits described in the text. The dashed lines indicate the fitted background component.}
\end{figure*}

Due to the long lifetime of the \PKzS\ and \PgL\ particles, the significance of the $\mathrm{V}^{0}$ decay length, which is the three-dimensional
distance between the primary and $\mathrm{V}^{0}$ vertices divided by its uncertainty,
is required to exceed 5. To remove \PKzS\ candidates misidentified as \PgL\ particles and vice
versa, the \PgL\ (\PKzS) candidate mass assuming both tracks to be pions (the lower-momentum track to be a pion and the higher-momentum track a proton) must differ by more than 20\,(10)\MeV from the nominal~\cite{PDG} \PKzS\,(\PgL) mass value. To remove photon conversions to an electron-positron pair, the mass of a \PKzS\ or \PgL\ candidate assuming both tracks to have the electron mass must exceed 15\MeV.
The angle $\theta^{\text{point}}$ between the $\mathrm{V}^{0}$ momentum vector and
the vector connecting the primary and $\mathrm{V}^{0}$ vertices
is required to satisfy $\cos\theta^{\text{point}}>0.999$. This reduces the contributions of particles from nuclear interactions, random combinations of tracks, and
secondary \PgL\ particles originating from the weak decays of
$\Xi$ and $\PgO$ particles.

To optimize the reconstruction of \PgXm\ particles, requirements on the three-dimensional impact parameter significance of its decay products with respect to the primary
vertex are applied. This significance must be larger than 3\,(4) for the proton (pion) tracks from the \PgL\ decay, and larger than 5 for the direct pion candidate from the \PgXm\ decay. To further reduce the background from random combinations of tracks, the corresponding impact parameter significance of \PgXm\ candidates cannot exceed 2.5. The three-dimensional decay length significance, with respect to the primary vertex, of the \PgXm\ candidate and the associated \PgL\ candidate must exceed 3 and 12, respectively.

The \PKzS, \PgL, and \PgXm\ reconstruction efficiencies are about 15,
5, and 0.7\% for $\pt \approx 1\GeV$, and 20, 10, and
2\% for $\pt>3\GeV$, averaged over $\abs{\eta}<2.4$. These efficiencies account for the effects of acceptance, and for the branching fractions of the decay modes in which the strange particles are reconstructed. The invariant mass distributions of reconstructed \PKzS, \PgL, and \PgXm\ candidates with $1<\pt<3\GeV$ are shown in Fig.~\ref{fig:v0mass}
for \pPb events with $220 \leq \noff < 260$. Prominent mass peaks are visible, with little background. The solid lines show the result of a maximum likelihood fit. In this fit, the strange-particle peaks are modeled as the sum of two Gaussian functions with a common mean. The ``average $\sigma$'' values in Fig.~\ref{fig:v0mass} are the square root of the weighted average of the variances of the two Gaussian functions. The background is modeled with a quadratic function for the \PKzS\ results, with the analytic form $Aq^{1/2}+Bq^{3/2}$ with $q = m - (m_{\Pgp} + m_{\Pp})$ for the \PgL\ results, and with the form $Cq^{D}$ with $q = m - (m_{\PgL} + m_{\Pgp})$ for the \PgXm\ results, where $A$, $B$, $C$, and $D$ are fitted parameters. These fit functions are found to provide a good description of the signal and background with relatively few free parameters. The fits are performed over the ranges of strange-particle invariant masses indicated in Fig.~\ref{fig:v0mass} to obtain the raw strange-particle yields $N^\text{raw}_{\PKzS/\PgL/\PgXm}$.

The raw strange-particle yields are corrected to account for the branching fraction of the reconstructed decay mode, and for the acceptance and reconstruction efficiency of the strange particle, using simulated event samples based on the \PYTHIA6 (\pp) or \EPOS\ (\pPb and \PbPb) event generator and \GEANTfour modeling of the detector:
\begin{linenomath}
\begin{equation}
\label{eq:efficiency}
N^\text{corr}_{\PKzS/\PgL/\PgXm} = \frac{N^\text{raw}_{\PKzS/\PgL/\PgXm}}{R_\text{corr}},
\end{equation}
\end{linenomath}
where $R_\text{corr}$ is a correction factor from simulation given by the ratio of the raw reconstructed yield to the total generated yield for the respective strange particle, with $N^\text{corr}_{\PKzS/\PgL/\PgXm}$ the corrected yield.

The raw \PgL\ particle yield includes contributions from the decays of \PgXm\ and \PgO\ particles.
This ``nonprompt'' contribution is largely determined by the relative \PgXm\ to \PgL\ yield (because the contribution from $\PgO$ particles is negligible). The stringent requirements placed on $\cos\theta^{\text{point}}$ remove a large fraction of the nonprompt \PgL\ component but, from simulation, up to 10\% of the \PgL\ candidates at high \pt are nonprompt. If the relative \PgXm\ to \PgL\ yield in simulation is modeled precisely, the contamination from nonprompt \PgL\ particles will be removed by the correction procedure of Eq. (\ref{eq:efficiency}). Otherwise, an additional correction to account for the residual contamination is necessary.
As the \PgXm\ particle yields are explicitly measured in this analysis, this residual correction
factor can be determined directly from the data as:
\begin{linenomath}
\begin{equation}
\label{PromptEfficiency}
f^\text{residual}_{\PgL, \mathrm{np}} = 1+f^{\text{raw},\mathrm{MC}}_{\PgL,\mathrm{np}} \left ( \frac{N^\text{corr}_{\PgXm}/N^\text{corr}_{\PgL}}{N^\mathrm{MC}_{\PgXm}/N^\mathrm{MC}_{\PgL}}-1 \right ),
\end{equation}
\end{linenomath}
where $f^{\text{raw}, \mathrm{MC}}_{\PgL,\mathrm{np}}$ denotes the fraction of nonprompt \PgL\ particles
in the raw reconstructed \PgL\ sample as determined from simulation, while
$N^\text{corr}_{\PgXm}/N^\text{corr}_{\PgL}$ and $N^\mathrm{MC}_{\PgXm}/N^\mathrm{MC}_{\PgL}$
are the \PgXm-to-\PgL\ yield ratios from the data after applying the corrections of Eq.~(\ref{eq:efficiency}),
and from generator-level simulation, respectively. The final prompt \PgL\
particle yield is given by $N^\text{corr}_{\PgL}/f^\text{residual}_{\PgL, \mathrm{np}}$.
Based on \EPOS\ MC studies, which has a similar \PgXm/\PgL\ ratio to the data, the residual nonprompt contributions to the \PgL\ yields are found to be negligible in \pPb and \PbPb collisions, while in \pp  collisions the correction is 1--3\% depending on the \pt value of the \PgL\ particle. Note that $N^\text{corr}_{\PgL}$ in Eq.~(\ref{PromptEfficiency}) is derived using Eq.~(\ref{eq:efficiency}), which in principle contains the residual nonprompt \PgL\ contributions. Nonetheless, by applying Eq.~(\ref{PromptEfficiency}) in an iterative fashion, we expect $N^\text{corr}_{\PgL}$ to approach a result corresponding to prompt \PgL\ particles only. A second
iteration of correction is found to have an effect of less than 0.1\% on the \PgL\ particle yield. As a cross-check we treat the sample of simulated events generated with the HIJING program like data and verify that we obtain the correct yields at the generator level after applying the correction procedure described above.

\section{Systematic uncertainties}

Table~\ref{tab:syst-table-Mid} summarizes the different sources of systematic uncertainty in the yields of each strange particle species. The values in parentheses correspond to the systematic uncertainties in the forward rapidity regions ($-2.4 < y_\text{cm} < -1.5$ and $0.8 < y_\text{cm} < 1.5$) for \pPb data, if they differ from those at mid-rapidity. The dominant sources of systematic uncertainty are associated with the strange-particle reconstruction, especially the efficiency determination.

\begin{table*}[ht]
\topcaption{\label{tab:syst-table-Mid} Summary of systematic uncertainties for the \pt spectra of \PKzS, \PgL, and \PgXm\ particles
  in the center-of-mass rapidity range $\abs{y_\text{cm}} < 1.0$ (for \pPb events, at forward rapidities, if different) for the three collision systems.}
\centering
\begin{tabular}{l*{5}{c}}
\hline
Source         & \multicolumn{2}{c}{\PKzS\ (\%)} & \multicolumn{2}{c}{\PgL\ (\%)} & {\PgXm\ (\%)}\\
\hline
\pt(\GeVns)               &${<}1.5$ &${>}1.5$ &${<}1.5$ & ${>}1.5$ & \\
\hline
Single-track efficiency &7.8 &7.8 &7.8 & 7.8 &11.7\\
Yield extraction &2 (3) &2 (3) &2 (4) &2 (4) &3\\
Selection criteria &3.6 (3.6) &2.2 (3.6) &3.6 (6.4) &2.2 (6.4) &7 \\
Momentum resolution &2 &2 &2 &2 &2\\
Nonprompt \PgL\ correction & & &2 &2 & \\
Pileup (\pp  only) &3 &1 &3 &1 &3\\
Proton direction (\pPb only) &3 (3) &3 (3) &3 (5) &3 (5) &4\\
Rapidity binning &1 (2) &1 (2) &1 (3) &1 (3) &2\\
Efficiency correction & & & & &5\\
\hline
Total (\pp) &9.6 &8.7 &9.8 &8.9 &15.4\\
Total (\pPb)  & 9.6 (10.0) &9.2 (10.0) & 9.8 (12.6) &9.4 (12.6) &15.6\\
Total (\PbPb) &9.1 &8.6 &9.3 &8.9 &15.1\\
\hline
\end{tabular}
\end{table*}

The systematic uncertainty in determining the efficiency of a single track is
3.9\%~\cite{CMS-PAS-TRK-10-002}. The tracking efficiency is strongly correlated
with the lifetime of a particle because when and where a particle decays determine how efficiently
the detector captures its decay products. We observe agreement of the \PKzS\ lifetime distribution
($c\tau$) between data and simulation, and similarly for the \PgL\ and \PgXm,
which provides a cross-check of the systematic uncertainty.
This translates into a systematic uncertainty in the reconstruction efficiency
of 7.8\% for the \PKzS\ and \PgL\ particles, and 11.7\% for the \PgXm\ particles. Different background fit functions and methods to extract the yields for the \PKzS, \PgL, and \PgXm\ are compared. The background fit function is varied to a fourth-order polynomial for the \PKzS\ and \PgL\ studies, and to a linear function for the \PgXm\ study. The yields are obtained by integrating over a region that is $\pm$5 times the average resolution and centered at the mean, rather than over the entire fitted mass range. Possible contamination by residual misidentified $\mathrm{V}^{0}$ candidates (\ie, a \PKzS\ particle misidentified as a \PgL\ particle, or vice versa) is investigated by varying the invariant mass range used
to reject misidentified $\mathrm{V}^{0}$ candidates. On the basis of these studies we assign systematic uncertainties of 2--4\% to the yields.
Systematic effects related to the selection of the strange-particle candidates are
evaluated by varying the selection criteria, resulting in
an uncertainty of 1--7\%. The impact of finite momentum resolution on the spectra is estimated using the \EPOS\ event generator. Specifically, the generator-level \pt spectra of the strange particles are smeared by the momentum resolution, which is determined through comparison of the generator-level and matched reconstructed-level particle information. The difference between the smeared and original spectra is less than 2\%. The systematic uncertainty associated with nonprompt \PgL\ corrections to the \PgL\ spectra is evaluated through propagation of the systematic uncertainty in the $N^\text{corr}_{\PgXm}/N^\text{corr}_{\PgL}$ ratio in Eq. (\ref{PromptEfficiency}) to the $f^\text{residual}_{\PgL, \mathrm{np}}$ factor, and is found to be less than 2\%. Systematic uncertainties
introduced by possible residual pileup effects for \pp  data are estimated to be 1--3\%. This uncertainty is evaluated through both tightening (only one reconstructed vertex allowed per event) and loosening (no event rejection on the basis of the number of vertices) the pileup rejection criteria~\cite{Chatrchyan:2013nka}.
The uncertainty associated with pileup is negligible for the \pPb and \PbPb data since there are very few events in those samples with more than one reconstructed vertex. In \pPb collisions, the direction of the p and Pb beams were reversed during course of the data collection, as mentioned in Section~\ref{sec:detector}. Comparison of the particle \pt spectra with and without the beam reversal yields an uncertainty of 2--5\% for all particle types. The effect of the choice of the rapidity bins is assessed by dividing each bin into two, thereby doubling the number of bins, resulting in a systematic uncertainty of 1--3\% for the \pt spectra. For the \PgXm, the reconstruction efficiency correction is smoothed by averaging adjacent bins in order to compensate for the limited statistical precision of the MC sample. Variations in the smoothing procedure lead to a systematic uncertainty of 5\% for the \pt spectra of the \PgXm.

All sources of systematic uncertainty are uncorrelated and summed in quadrature to define the total systematic uncertainties in the \pt spectra of each strange particle. The total systematic uncertainties between the \pp, \pPb, and \PbPb\ systems are similar and largely correlated. When calculating ratios of particle yields, most of the systematic uncertainties partially or entirely cancel.  For example, the systematic uncertainties due to tracking efficiency and pileup for the \PgL/2\PKzS\ ratio are negligible.

\section{Results}
\subsection{Multiplicity dependence at mid-rapidity}
\label{subsec:res_mid}

\begin{figure*}
\centering
\includegraphics[width=\linewidth]{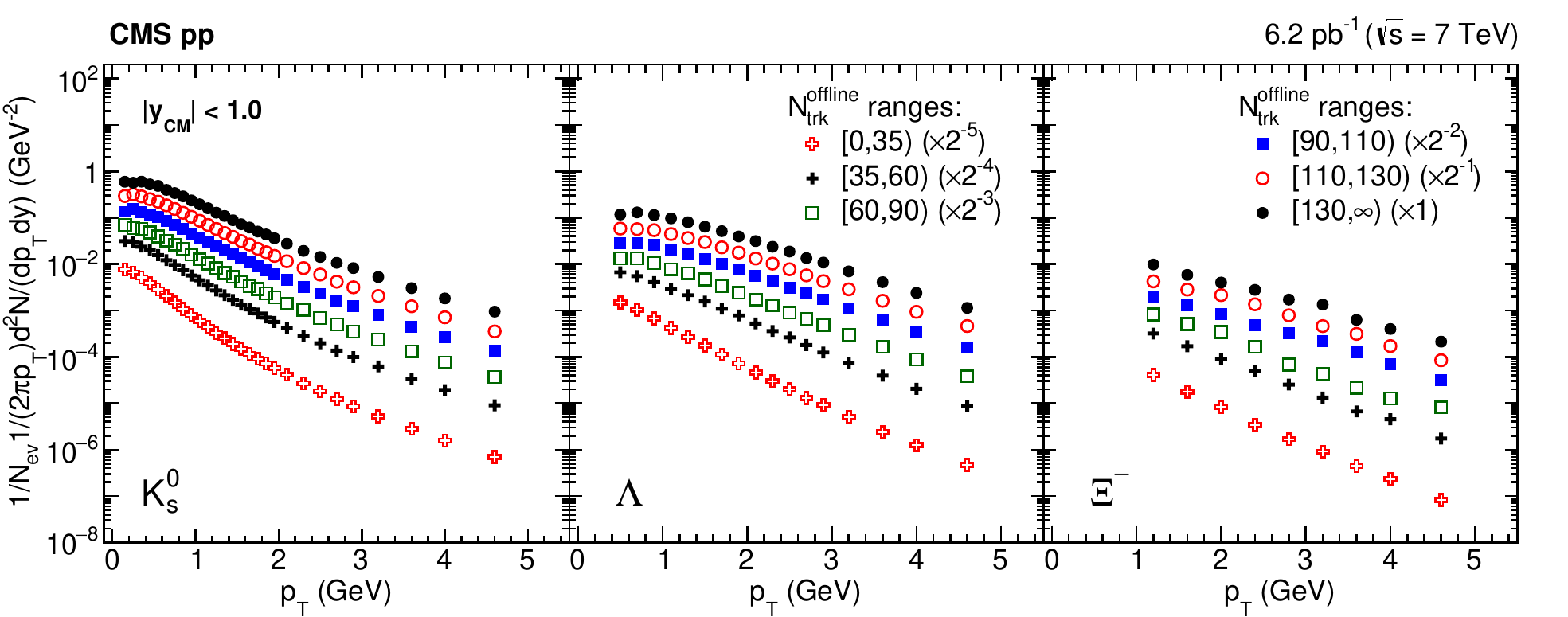}
\includegraphics[width=\linewidth]{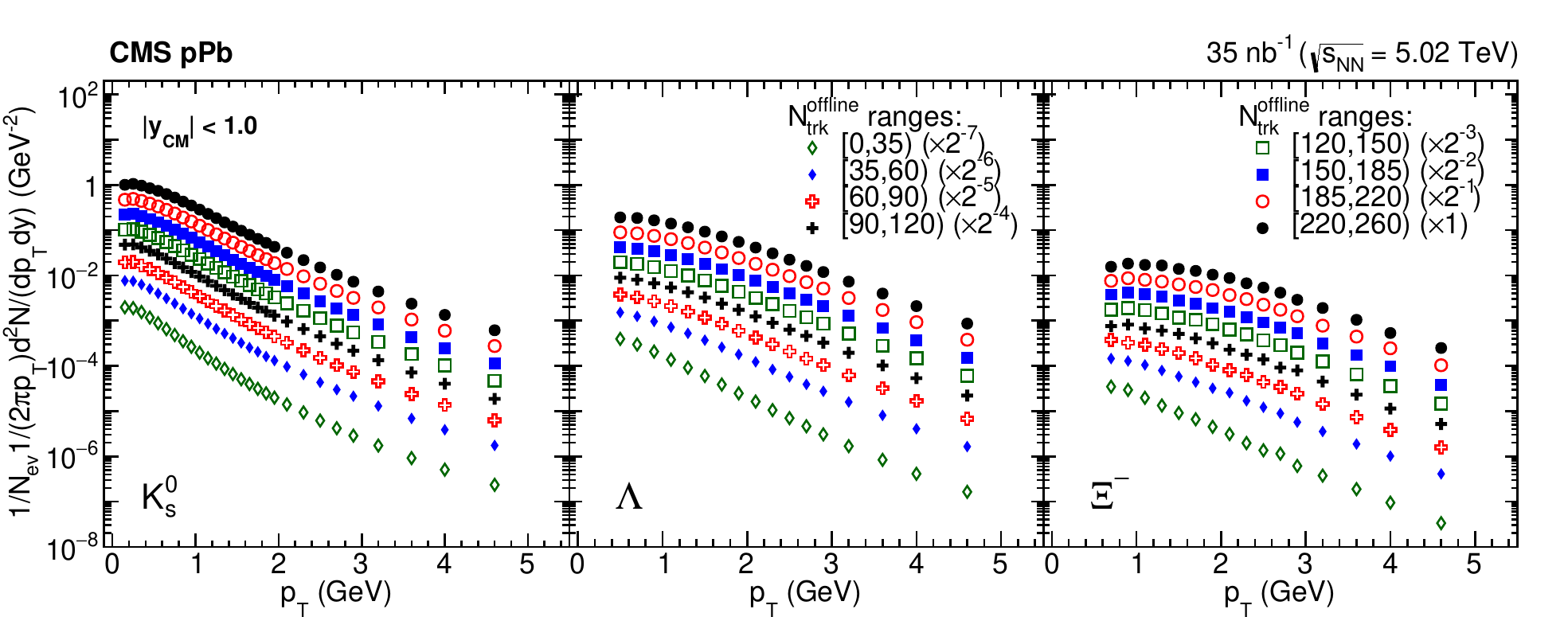}
\includegraphics[width=\linewidth]{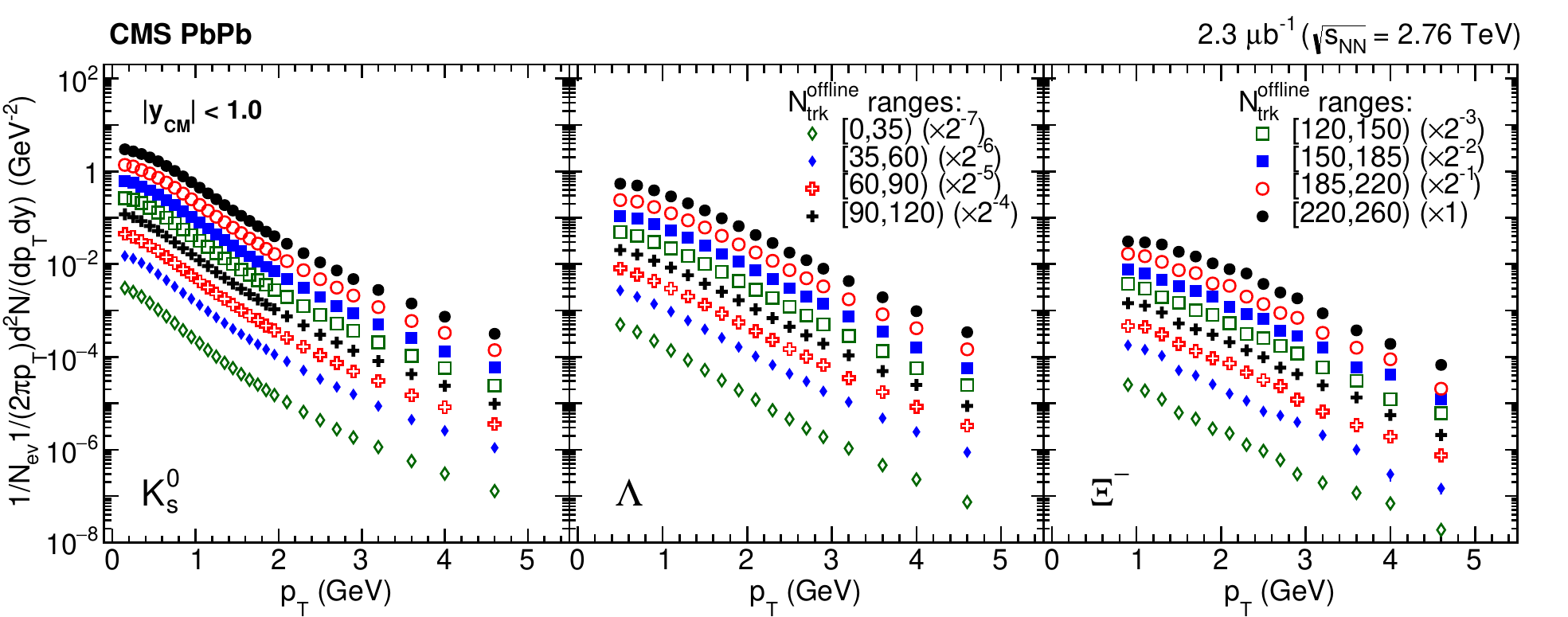}
  \caption{ \label{fig:V0spectra_midRapidity}
        The \pt spectra of \PKzS, \PgL, and \PgXm\ particles in the center-of-mass
  	rapidity range $\abs{y_\text{cm}} < 1$ in \pp  collisions at $\roots = 7\TeV$\ (top),
    \pPb collisions at $\roots = 5.02\TeV$\ (middle), and \PbPb collisions at $\rootsNN = 2.76\TeV$\ (bottom)
    for different multiplicity intervals. The inclusion of the charge-conjugate states is implied for \PgL\ and \PgXm\ particles. The data in the different multiplicity intervals are scaled by factors of $2^{-n}$ for better visibility. The statistical uncertainties are smaller than the markers and the systematic uncertainties are not shown.
}
\end{figure*}

The $\pt$ spectra of \PKzS, \PgL, and \PgXm\ particles
with $\abs{y_\text{cm}}<1$ in \pp  collisions at $\roots = 7\TeV$ (top), \pPb collisions at $\roots = 5.02\TeV$ (middle), and \PbPb collisions
at $\rootsNN = 2.76\TeV$\ (bottom) are presented in Fig.~\ref{fig:V0spectra_midRapidity},
for different multiplicity intervals. Due to details in the implementation of the dedicated
high-multiplicity trigger thresholds used to select the \pp  events, the multiplicity intervals for \pp  events differ slightly from those for \pPb and \PbPb events. The $\pt$ differential yield is defined as $\rd N^{2}/(2\pi \pt)\rd\pt\,\rd y$. For the purpose of better visibility, the data are scaled by factors of $2^{-n}$, as indicated in the figure legend.
A clear evolution of the spectral shape with multiplicity can be seen for
each particle species in each collision system. For higher multiplicity events,
the spectra tend to become flatter (i.e., "harder"), indicating a larger
$\KETavg$ value. Within each collision system,
heavier particles (\eg, \PgXm) exhibit a harder spectrum than lighter particles (\PKzS),
especially for high-multiplicity events.

\begin{figure*}[thb]
\centering
\includegraphics[width=\linewidth]{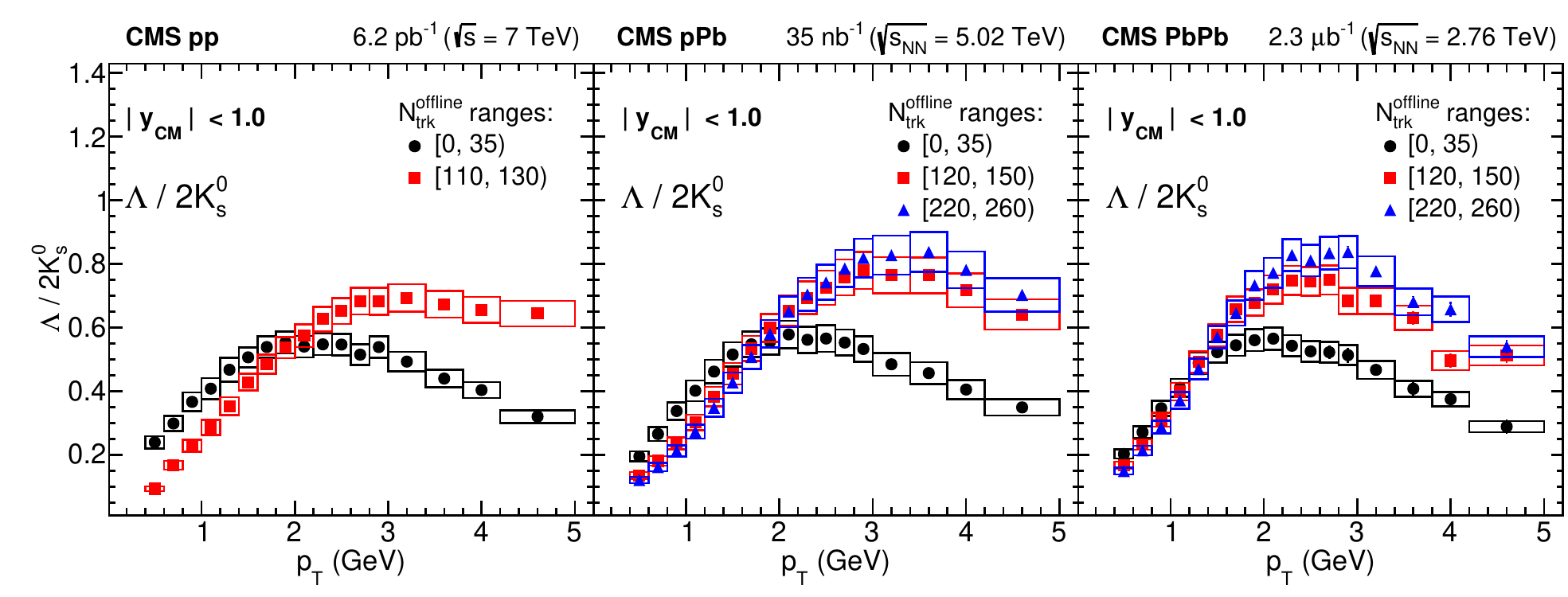}
\includegraphics[width=\linewidth]{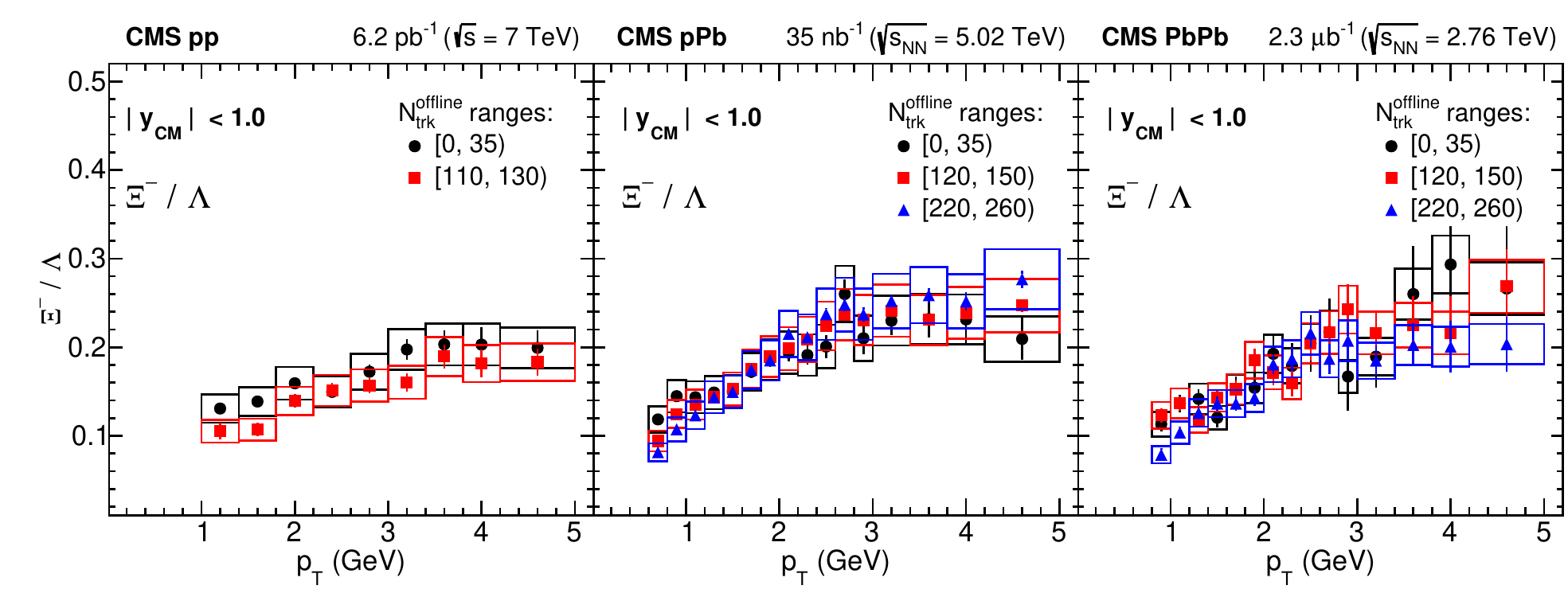}
  \caption{ \label{fig:ParticlesRatio} Ratios of \pt spectra for \PgL/2\PKzS\ (top) and \PgXm/\PgL\ (bottom)
  in the center-of-mass rapidity range $\abs{y_\text{cm}} < 1.0$ for \pp  collisions at $\roots = 7\TeV$\ (left),
  \pPb collisions at $\roots = 5.02\TeV$\ (middle), and \PbPb collisions at $\rootsNN = 2.76\TeV$\ (right).
  Two (for \pp) or three (for \pPb and \PbPb) representative multiplicity intervals are presented. The inclusion of the charge-conjugate states is implied for \PgL\ and \PgXm\ particles. The error bars represent the statistical
  uncertainties, while the boxes indicate the systematic uncertainties.
}
\end{figure*}

To examine the differences in the multiplicity dependence of the spectra in greater detail, the ratios \PgL/2\PKzS\ and \PgXm/\PgL\ of the yields are shown in Fig.~\ref{fig:ParticlesRatio} as a function of \pt for different multiplicity ranges
in the \pp, \pPb, and \PbPb systems. The results for the \PgL/2\PKzS\ ratio are shown in Fig.~\ref{fig:ParticlesRatio} (top). For $\pt \lesssim 2\GeV$, the \PgL/2\PKzS\ ratio is seen to be smaller in high-multiplicity events than in low-multiplicity events for a given \pt value. In \pp  and \pPb collisions,
this trend is similar to what has been observed between peripheral and
central \PbPb collisions~\cite{Abelev:2013vea}; this trend is not as evident
for the \PbPb data in Fig.~\ref{fig:ParticlesRatio} (top right) because in the present study only
\PbPb events of 50--100\% centrality are considered. At higher \pt, this multiplicity ordering of the \PgL/2\PKzS\ ratio is reversed. In hydrodynamic models such as those presented in Refs.~\cite{Bozek:2012qs,Karpenko:2012yf}, this behavior can be interpreted as the effect of radial flow.
A stronger radial flow is developed in higher-multiplicity events, which boosts heavier particles (\eg, \PgL) to higher \pt, resulting in a suppression of the \PgL/2\PKzS\ ratio at low \pt.
Comparing the various collision systems at low \pt, the difference
in the \PgL/2\PKzS\ ratio between low- and high-multiplicity events is seen to be largest for the \pp  data. In the hydrodynamic model of Ref.~\cite{Shuryak:2013ke}, smaller collision systems like \pp  produce a larger radial-flow effect than larger systems like \pPb or \PbPb, for similar multiplicities, which could explain this observation. For $\pt\ > 2\GeV$, the baryon enhancement could be explained by recombination models, in which free quarks recombine to form hadrons~\cite{Fries:2008hs}. In previous studies (e.g., Ref.~\cite{Chatrchyan:2012qb}), it has been shown that the average \pt value of various particle species has only a slight center-of-mass energy dependence (10\% at high multiplicity). This dependence is not sufficient to explain the differences observed in Fig.~\ref{fig:ParticlesRatio} between the various systems.

For each multiplicity interval, the \PgL/2\PKzS\ ratio reaches a
maximum that has a similar value for all three collision processes, and then decreases at higher \pt. The location of the maximum increases with multiplicity from around $\pt = 2$ to 3\GeV.

The results for the \PgXm/\PgL\ ratio are shown in Fig.~\ref{fig:ParticlesRatio} (bottom). In this case, the difference between the low-
and high-multiplicity events is much smaller than for the \PgL/2\PKzS\ ratio, for all three collisions systems. For all systems, the \PgXm/\PgL\ ratio increases with \pt and reaches a plateau at around $\pt = 3\GeV$. Due to the large systematic uncertainty, it is not possible to draw a conclusion with respect to the radial-flow interpretation.

\begin{figure}[thb]
\centering
\includegraphics[width=\cmsFigWidth]{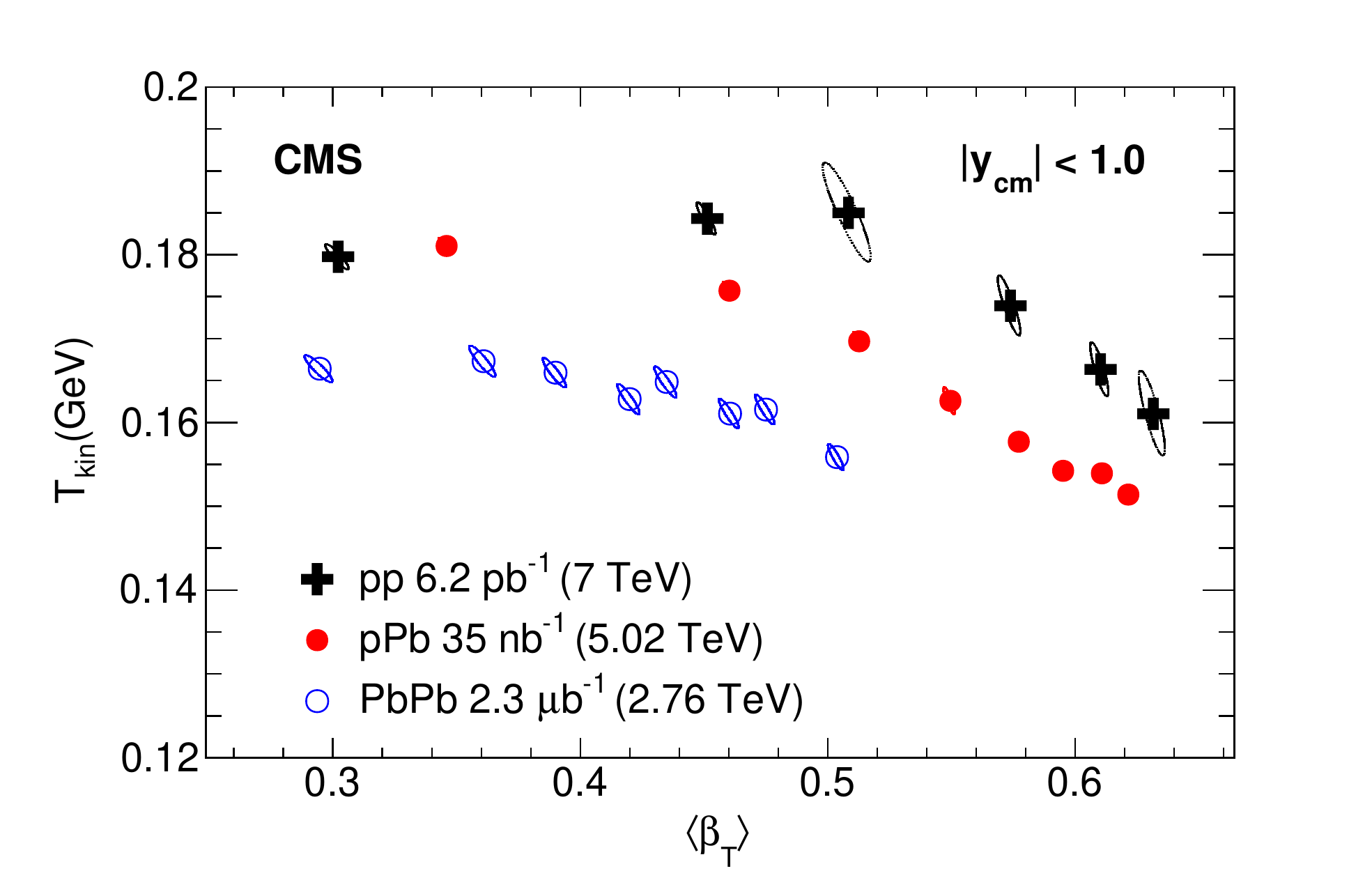}
  \caption{ \label{fig:BlastWaveFits_midrapidity}
  The extracted kinetic freeze-out temperature, $T_\text{kin}$, versus the average
  radial-flow velocity, $\langle \beta_\mathrm{T} \rangle$, from a simultaneous
  blast-wave fit to the \PKzS\ and \PgL\ \pt spectra at $\abs{y_\text{cm}} < 1$ for different
  multiplicity intervals in \pp, \pPb, and \PbPb collisions. The six \pp  and eight \pPb and \PbPb   multiplicity intervals are indicated in the legend of Fig.~\ref{fig:V0spectra_midRapidity}. For the results in this plot, the multiplicity increases from left to right. The correlation
  ellipses represent the statistical uncertainties. Systematic uncertainties, which are evaluated to be on the order of a few percent, are not shown.
}
\end{figure}

Motivated by the hydrodynamic model, we perform a simultaneous fit of a blast-wave function~\cite{Schnedermann_blastwave} to the \PKzS\ and \PgL\ spectra in Fig.~\ref{fig:V0spectra_midRapidity}. The fits are restricted to low \pt because that is the region in which the blast-wave model is valid. The blast-wave model is strictly appropriate only for directly produced particles, while about 1/3 of the \PKzS\ mesons may be from higher mass resonances~\cite{Adam:2016bpr}. The \PgXm\ particle is not used in the fit as there are not many \PgXm\ at low \pt. The fits are performed for each collision system separately. The fit ranges are $0.1 < \pt < 1.5\GeV$ for the \PKzS\ and $0.6 < \pt < 3.0\GeV$ for the \PgL. The fitted function is:
\begin{linenomath}
\begin{equation}
\label{blastwaveFunctionalForm}
\frac{1}{\pt}\frac{\mathrm{\rd N} }{\rd \pt}\sim \int_{0}^{R} r\, \rd r\, m_\mathrm{T} I_{0}\left ( \frac{\pt \sinh{\rho}}{T_\text{kin}} \right )K_{1}\left ( \frac{m_\mathrm{T} \cosh{\rho}}{T_\text{kin}} \right ),
\end{equation}
\end{linenomath}
where $\rho = \tanh^{-1}{\beta_\mathrm{T}} = \tanh^{-1}\left ( \beta_\mathrm{s}( {r}/{R})^{n} \right )$ is the velocity profile, $R$ is the radius of the medium (set to unity in the fit), $r$ is the radial distance from the center of the medium in the transverse plane, $n$ is the exponent of the velocity profile, $\beta^{}_\mathrm{T}$ is the transverse expansion velocity (also known as the radial-flow velocity), $\beta^{}_\mathrm{s}$ is the transverse expansion velocity on the surface of the medium, $T_\text{kin}$ is the kinetic freeze-out temperature, and $I_{0}$ and $K_{1}$ are modified Bessel functions. The fitted parameters that govern the shape are $n$, $\beta^{}_\mathrm{s}$, and $T_\text{kin}$.

In the blast-wave model, common values of $T_\text{kin}$ and average radial-flow velocity $\langle \beta_\mathrm{T} \rangle$ are assumed for all particle species, as is expected if the system is locally thermalized and undergoes a radial-flow expansion.
It is useful to directly compare the extracted values of $T_\text{kin}$ and $\langle \beta_\mathrm{T} \rangle$ from the different systems to study the system-size dependence at similar multiplicities.

\begin{figure*}[thb]
\centering
\includegraphics[width=0.7\textwidth]{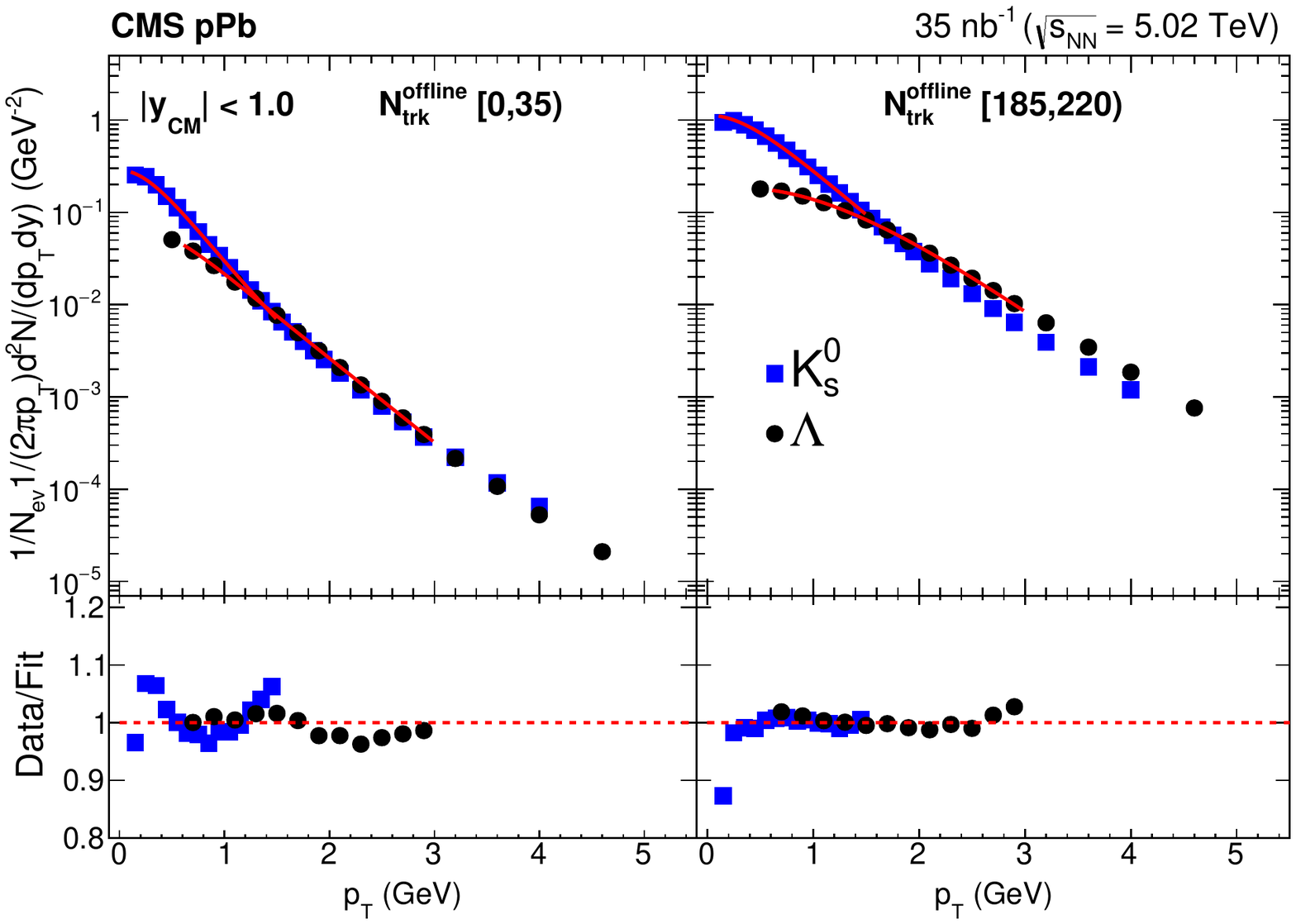}
  \caption{ \label{fig:pPb_blastWaveFitter_qualityPlots_2}
    Examples of simultaneous blast-wave fits of the \pt spectra of \PKzS\ and \PgL\ particles in low- and high-multiplicity \pPb events. The inclusion of the charge-conjugate states is implied for \PgL\ particles. The ratios of the fits to the data as a function of \pt are shown in the bottom panels.
    The uncertainties are statistical only and are too small to be visible for most of the points.
}
\end{figure*}

The extracted values of $T_\text{kin}$ and $\langle \beta_\mathrm{T} \rangle$ are shown in Fig.~\ref{fig:BlastWaveFits_midrapidity} for the six \pp  and for the eight \pPb and \PbPb multiplicity intervals. In this figure, the multiplicity increases from left to right. The ellipses correspond to one standard deviation statistical uncertainties, which for \pp\ collisions are smaller at low and high multiplicity due to the use of events collected with minimum bias and high-multiplicity triggers. Systematic uncertainties, which are evaluated by propagating the systematic uncertainties from the spectra to the blast-wave fits and altering the fit ranges, are on the order of a few percent and are not shown.
Examples of the fits are shown in Fig.~\ref{fig:pPb_blastWaveFitter_qualityPlots_2} for a low- and high-multiplicity range in \pPb collisions. In general, the
fit quality is good for high-multiplicity events except for the lowest
\pt range, while for low-multiplicity events there are discrepancies on
the order of 5\%. However, the discrepancies between the fit and data lie
within the systematic uncertainty.

The precise meaning of the $T_\text{kin}$ and $\langle \beta_\mathrm{T} \rangle$
parameters is model dependent, and they should not be interpreted literally as the kinetic
freeze-out temperature and radial-flow velocity of the system. The main purpose of
Fig.~\ref{fig:BlastWaveFits_midrapidity} is to provide a qualitative
comparison of the spectral shapes in the three systems. In the context of the blast-wave model, when comparing at similar multiplicities, the $T_\text{kin}$ parameter has the same value within 15\% among the three systems, while the $\langle \beta_\mathrm{T} \rangle$
parameter is larger when the system is smaller, i.e.,
$\langle \beta_\mathrm{T} \rangle_{\text \pp  } > \langle \beta_\mathrm{T} \rangle_{\text \pPb } > \langle \beta_\mathrm{T} \rangle_{\text \PbPb }$. This is qualitatively consistent
with the prediction of Ref.~\cite{Shuryak:2013ke}. The results of blast-wave
fits are known to depend on the particle species. Due to the limited
set of particles in this analysis, future studies will be needed
to further substantiate the conclusions.

\begin{figure*}[thb]
\centering
\includegraphics[width=\linewidth]{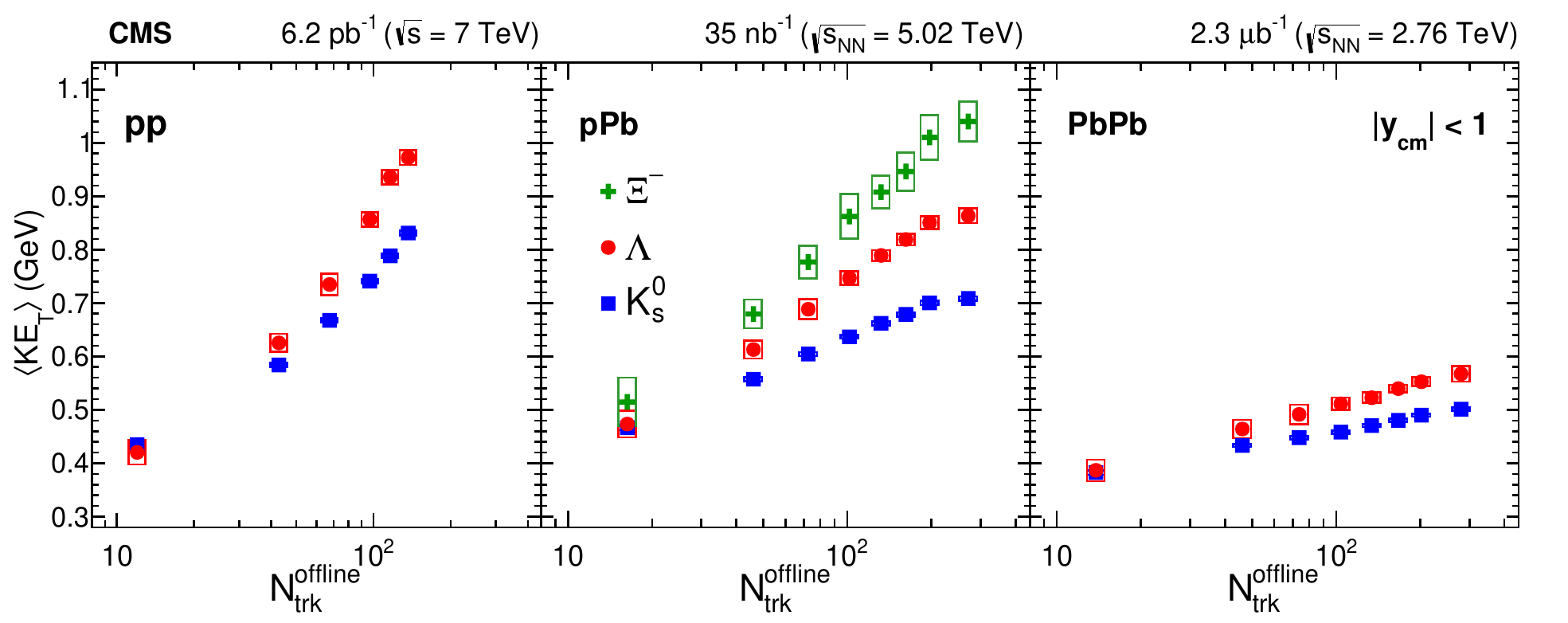}
  \caption{ \label{fig:MeanMt_pPb_midrapidity}
    The average transverse kinetic energy, $\KETavg$,
    at $\abs{y_\text{cm}}<1$ for \PKzS, \PgL, and \PgXm\  particles
    as a function of multiplicity in \pp, \pPb, and \PbPb collisions. The inclusion of the charge-conjugate states is implied for \PgL\ and \PgXm\ particles.
    For the \PgXm, only results from \pPb collisions are shown.
    The error bars represent the statistical uncertainties, while the boxes
    indicate the systematic uncertainties.
    }
\end{figure*}

The evolution of the \pt spectra with multiplicity can be compared more directly between the three systems through examination of the $\KETavg$ value.
The $\KETavg$ values at $\abs{y_\text{cm}}<1$ for \PKzS,
\PgL, and \PgXm\ particles as a function of multiplicity are shown in Fig.~\ref{fig:MeanMt_pPb_midrapidity}.
Extrapolation of the \pt spectra down to $\pt = 0$\GeV is a crucial step in
extracting the $\KETavg$ values, while the impact of the extrapolation up to $\pt\approx\infty$ is negligible, both on the value of $\KETavg$ and its uncertainty.
For the \PgXm\ particle, only results in \pPb collisions are shown due to the limitation of the low-\pt reach
in \pp  and \PbPb collisions, as can be seen from Fig.~\ref{fig:V0spectra_midRapidity}.
Blast-wave fits to the individual spectra, which only consider the spectrum shape but do not impose any physics constraint, are used to obtain the extrapolation.
The fraction of the extrapolated yield with respect to the total yield is about 1.2--2.5\% for the \PKzS, 5.8--15.1\% for the \PgL,
and 5.4--20.4\% for the \PgXm\ particles, depending on the multiplicity. Alternative methods to perform the extrapolation are used to evaluate a systematic uncertainty, including use of the predictions from the simultaneous blast-wave fit to the \PKzS\ and \PgL\ \pt spectra, and a linear extrapolation from the yields in a low range of \pt. The systematic uncertainties from Table~\ref{tab:syst-table-Mid} are also included in the evaluation of the $\KETavg$ uncertainties.

For the lowest multiplicity range, the $\KETavg$ values
for each particle species are seen to be similar. For all particle species, $\KETavg$ increases with increasing multiplicity.
However, the slope of the increase differs for different particles, with the heavier particles exhibiting a faster growth in $\KETavg$ for all systems. For a given multiplicity range, the $\KETavg$ value is roughly proportional to the particle's mass.
In PbPb collisions, this can be understood to be due to the onset of radial flow~\cite{STAR,PHENIX}. The observed difference between particle species at high multiplicity is seen to be larger for \pp  and \pPb events than for \PbPb events. Note, however, the difference in the center-of-mass energies between the three systems.

\begin{figure*}
\centering
\includegraphics[width=0.67\textwidth]{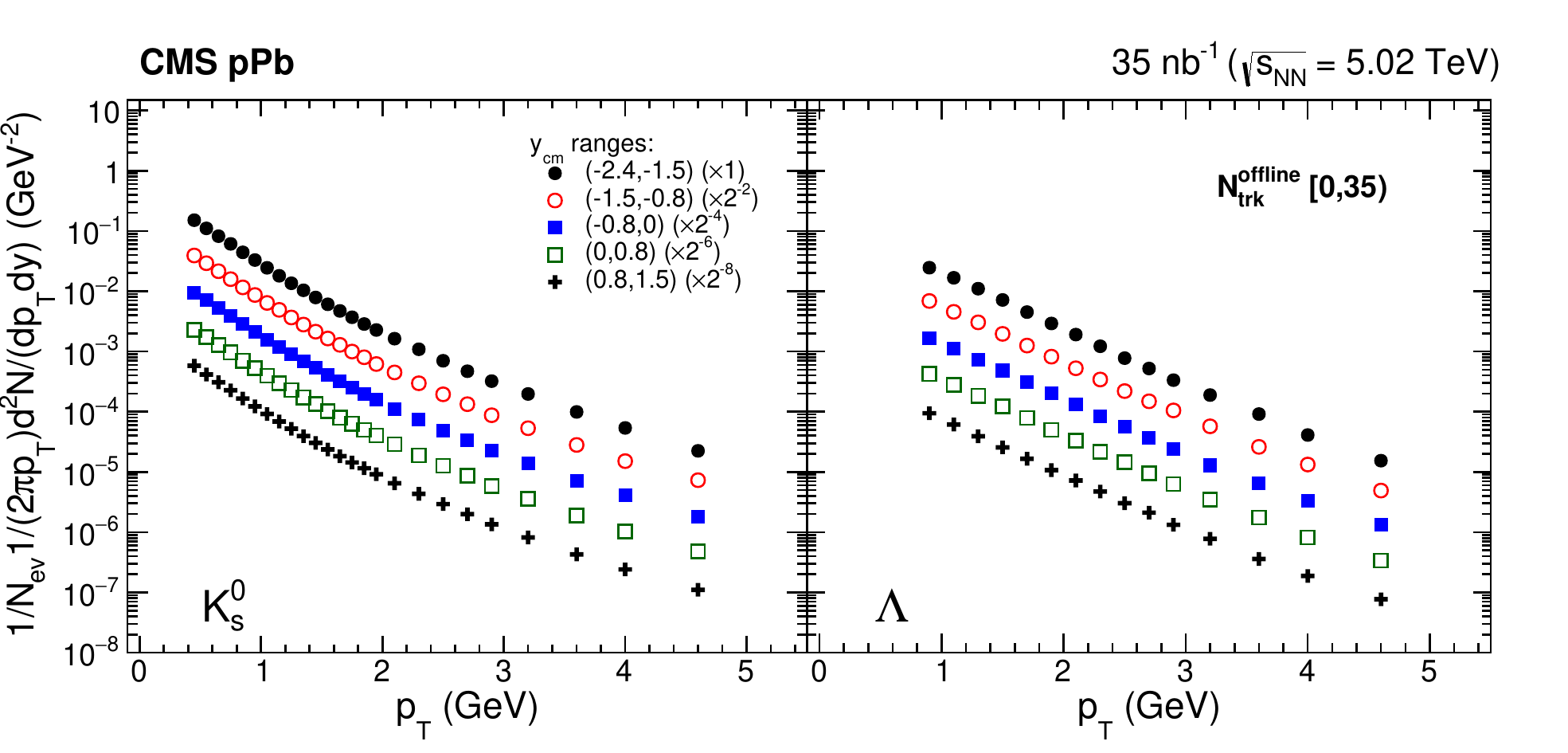}
\includegraphics[width=0.67\textwidth]{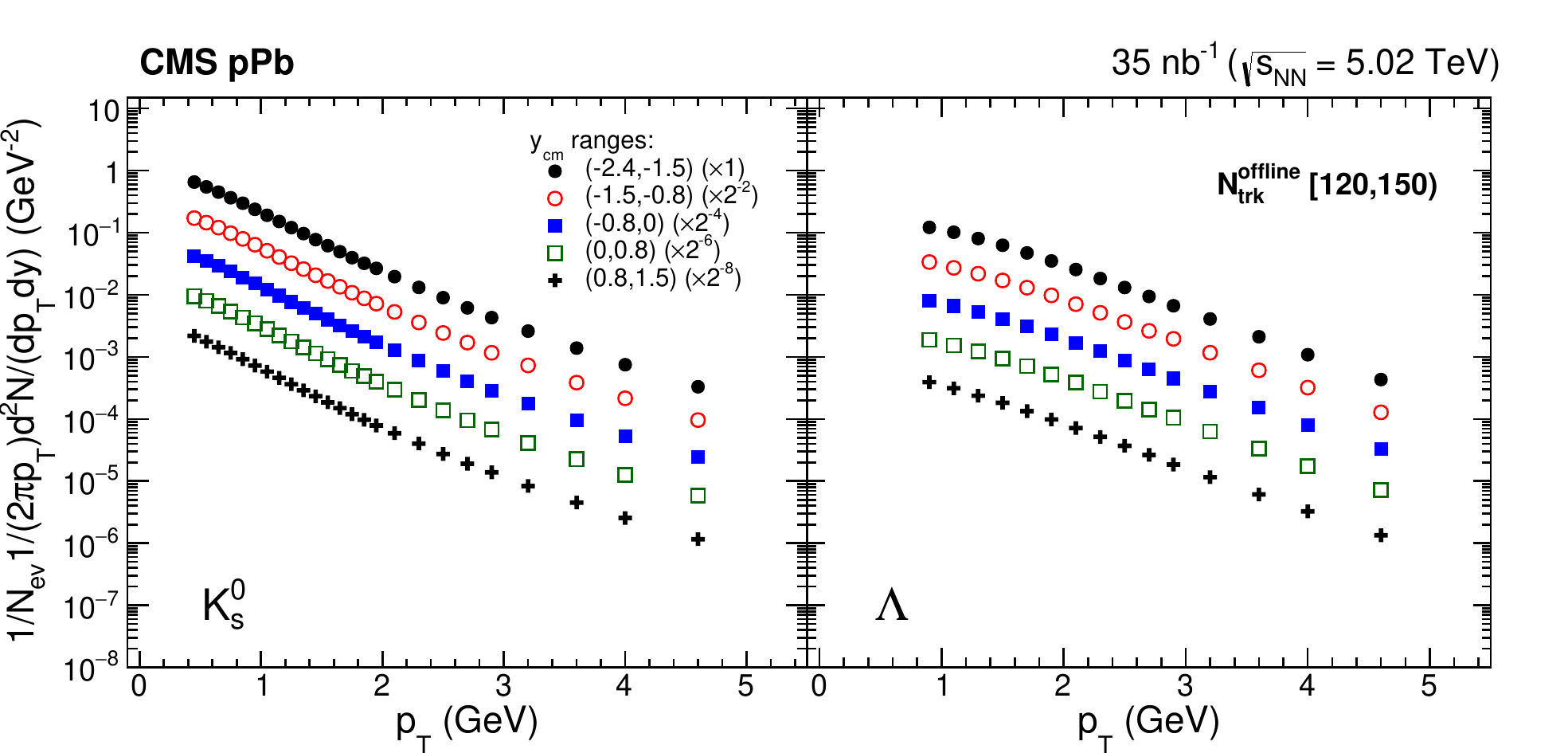}
\includegraphics[width=0.67\textwidth]{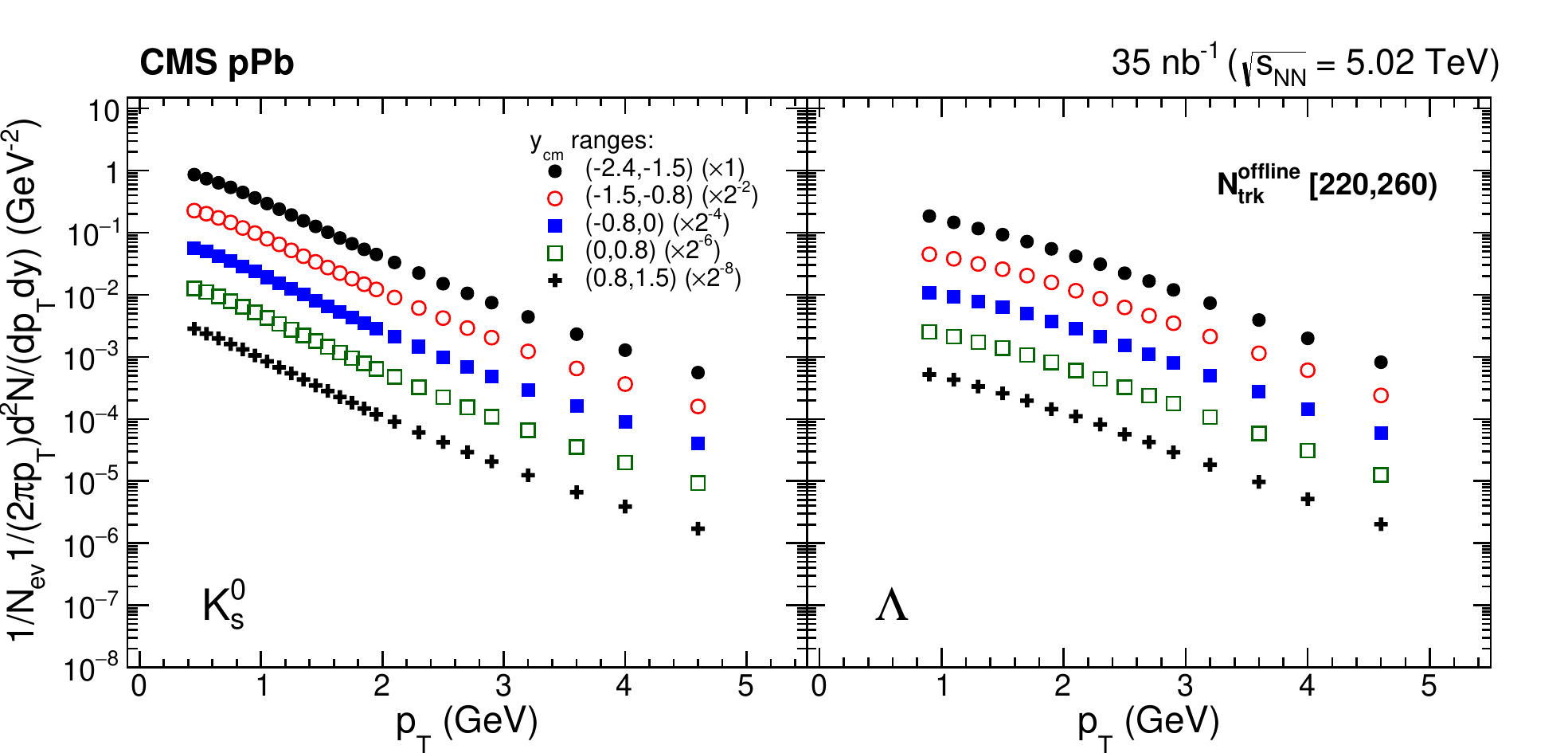}
  \caption{ \label{fig:V0spectra_rapidityDependence}
    The \pt spectra of \PKzS\ and \PgL\ particles in different $y_\text{cm}$ ranges for \pPb collisions at $\roots = 5.02\TeV$. The inclusion of the charge-conjugate states is implied for \PgL\ and particles.
    Results are shown for three multiplicity ranges: $0 \leq \noff<35$ (top), $120 \leq \noff<150$ (middle), and  $220 \leq \noff<260$ (bottom).
    Within each panel, the curves on top represent Pb-going events and the curves on bottom p-going events. The data in the different rapidity intervals are scaled by factors of $2^{-2n}$ for better visibility. The statistical uncertainties are smaller than the markers and the systematic uncertainties are not shown.
  }
\end{figure*}

\subsection{Rapidity dependence in \pPb events}

The rapidity dependence of the \pt spectra of the \PKzS\ and \PgL\ particles is studied in the \pPb data. No results for \PgXm\ particles are presented due to statistical limitations.
As a \pPb collision is asymmetric in rapidity, it is interesting to
compare the spectra along the Pb-going ($y_\text{cm}<0$) and p-going ($y_\text{cm}>0$) directions~\cite{Bozek:2013sda}.
The \pt spectra of \PKzS\ and \PgL\ particles
in different $y_\text{cm}$ ranges are shown in Fig.~\ref{fig:V0spectra_rapidityDependence} for small (top), intermediate (middle), and large (bottom) average multiplicities.
  	
\begin{figure*}[thb]
\centering
\includegraphics[width=0.67\textwidth]{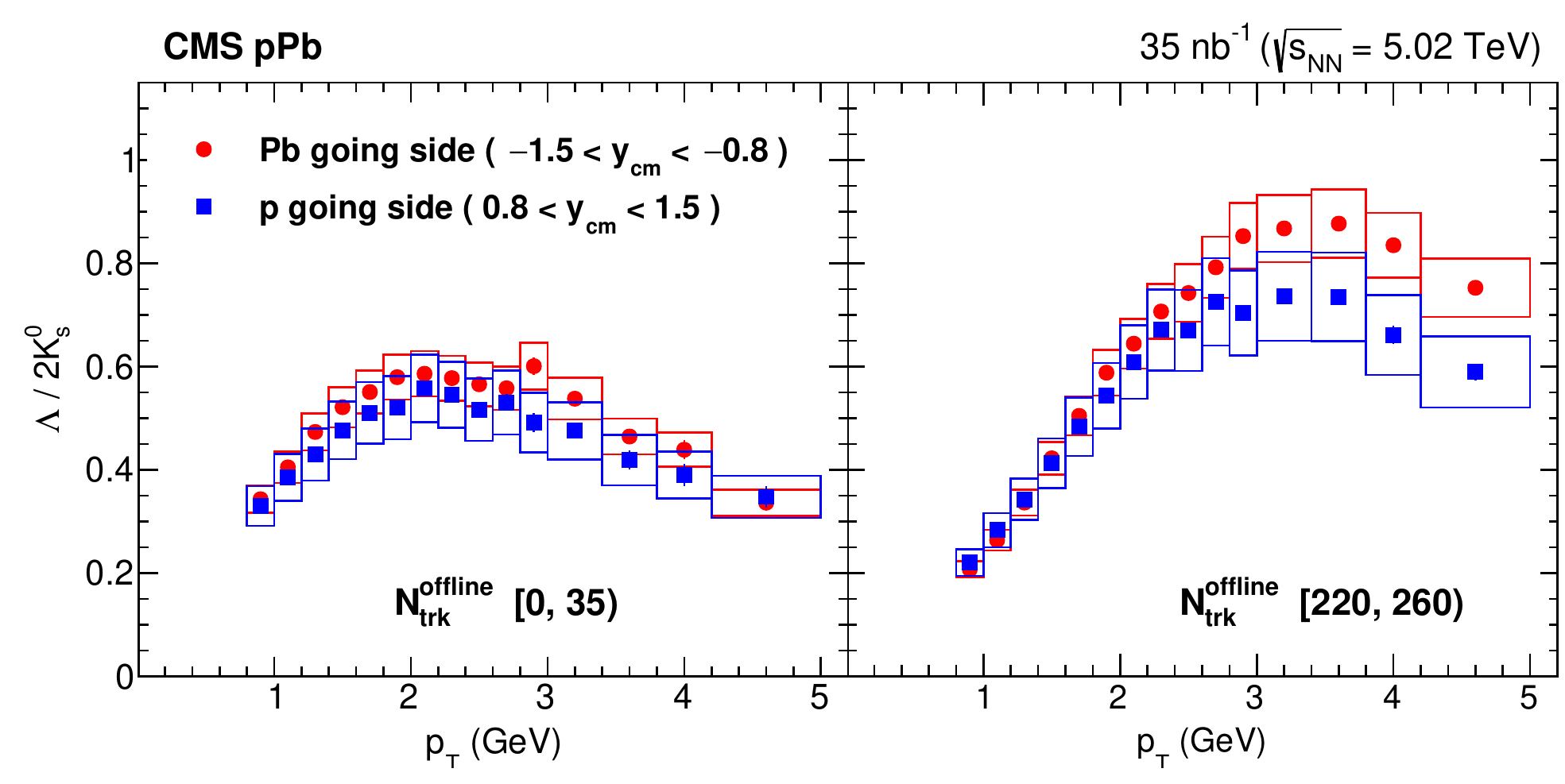}
  \caption{ \label{fig:V0spectra_rapidityRatio_highMultiplicity_to_lowMultiplicity}
  Ratios of \pt spectra, \PgL/2\PKzS, from the $-1.5 < y_\text{cm} < -0.8$ (Pb-going)
  and $0.8 < y_\text{cm} < 1.5$ (p-going) rapidity regions in \pPb collisions at $\roots = 5.02\TeV$. The inclusion of the charge-conjugate states is implied for \PgL\ particles.
  Results are presented for two multiplicity ranges $0 \leq \noff<35$ (left) and $220 \leq \noff<260$ (right).
  The error bars represent the statistical uncertainties, while the boxes indicate the systematic uncertainties.
}
\end{figure*}

The \PgL/2\PKzS\ ratios from the $-1.5 < y_\text{cm} < -0.8$ (Pb-going)
and $0.8 < y_\text{cm} < 1.5$ (p-going) rapidity regions are compared in Fig.~\ref{fig:V0spectra_rapidityRatio_highMultiplicity_to_lowMultiplicity}
for multiplicity ranges $0 \leq \noff < 35$ and $220 \leq \noff < 260$. For both the low-multiplicity and the high-multiplicity events, the \PgL/2\PKzS\ ratio from the Pb-going direction lies above the results from the p-going direction, with the largest difference observed at high \pt in the high-multiplicity sample.

\begin{figure*}[thb]
\centering
\includegraphics[width=0.67\textwidth]{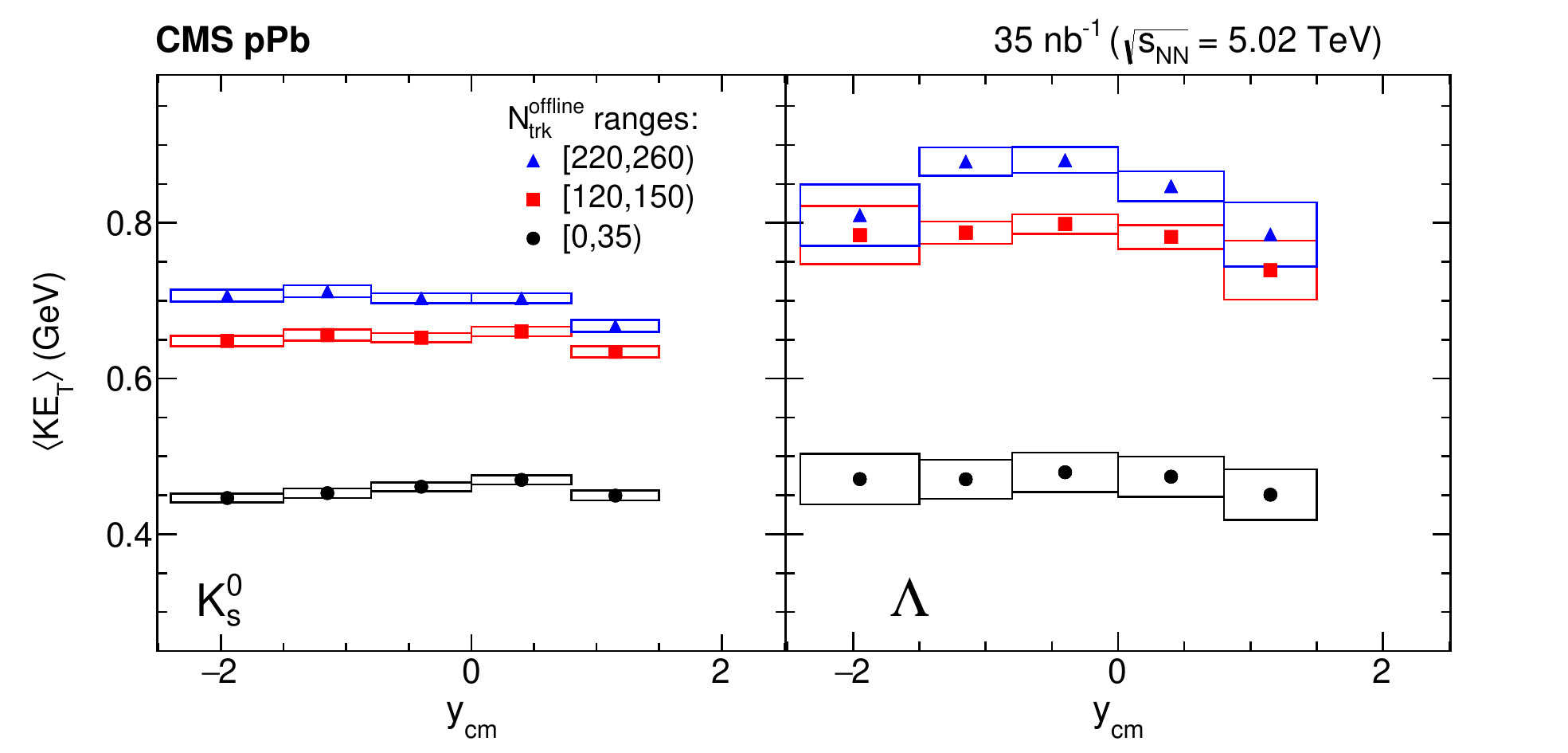}
  \caption{ \label{fig:K0sLam_meanPt_rapidityHist}
    The average transverse kinetic energy, $\KETavg$, as a function of $y_\text{cm}$ for the \PKzS\ and \PgL\ particles in three ranges of multiplicity in \pPb collisions at $\roots = 5.02\TeV$. The inclusion of the charge-conjugate states is implied for \PgL\ particles.
    The error bars represent the statistical uncertainties, while the boxes indicate the systematic uncertainties.  }
\end{figure*}

As a further study, we calculate $\KETavg$, following the procedure outlined in Section~\ref{subsec:res_mid}, and examine its dependence on $y_\text{cm}$ for \PKzS\ and \PgL\ particles in the \pPb collisions. The results are shown in Fig.~\ref{fig:K0sLam_meanPt_rapidityHist}. Although the systematic uncertainties at forward rapidities are large, the $\KETavg$ values are seen to become slightly asymmetric as multiplicity increases.
At low multiplicities ($0 \leq \noff<35$), the ratios of $\KETavg$ between the Pb-going side ($-1.5 < y_\text{cm} < -0.8$) and the p-going side ($0.8 < y_\text{cm} < 1.5$) are $1.01 \pm 0.01$ (syst.) for \PKzS\ particles and $1.04 \pm 0.05$ (syst.) for \PgL\ particles, both of which are consistent with unity within the systematic uncertainties (the statistical uncertainties are negligible).
However, in the highest multiplicity range, $220 \leq N_\text{trk}^\text{offline} < 260$, the ratios become
$1.06 \pm 0.01\syst$ for \PKzS\ particles and $1.12 \pm 0.06\syst$ for \PgL\ particles, suggesting that an asymmetry
in $\KETavg$ is developed between the Pb-going and p-going sides. This trend is qualitatively consistent with the hydrodynamic prediction for \pPb collisions~\cite{Bozek:2013sda}.

\section{Summary}
\label{sec:conclusion}

Measurements of strange hadron (\PKzS, \PgL+\PagL, and \PgXm+\PagXp) transverse momentum spectra
in \pp, \pPb, and \PbPb collisions are presented over a wide range of event charged-particle multiplicity
and particle rapidity. The study is based on samples of \pp  collisions at $\roots = 7\TeV$, \pPb collisions at $\roots = 5.02\TeV$, and \PbPb collisions at $\rootsNN = 2.76\TeV$, collected with the CMS detector at the LHC. In the context of hydrodynamic models, the measured particle spectra are fitted with a blast wave function, which describes an expanding fluid-like system. When comparing at a similar multiplicity, the extracted radial-flow velocity parameters are found to be larger in \pp  and \pPb collisions than that in \PbPb collisions. The average transverse kinetic energy $\KETavg$ of strange hadrons is observed to increase with multiplicity, with a stronger increase for heavier particles. At similar multiplicities, the difference in $\KETavg$ between the strange-particle species is larger in the smaller \pp  and \pPb systems than in the \PbPb system. For \pPb collisions, $\KETavg$ in the Pb-going direction for \PKzS\ (\PgL+\PagL) is 6\% (12\%) larger than in the p-going direction for events with the highest particle multiplicities.

\begin{acknowledgments}
We congratulate our colleagues in the CERN accelerator departments for the excellent performance of the LHC and thank the technical and administrative staffs at CERN and at other CMS institutes for their contributions to the success of the CMS effort. In addition, we gratefully acknowledge the computing centers and personnel of the Worldwide LHC Computing Grid for delivering so effectively the computing infrastructure essential to our analyses. Finally, we acknowledge the enduring support for the construction and operation of the LHC and the CMS detector provided by the following funding agencies: BMWFW and FWF (Austria); FNRS and FWO (Belgium); CNPq, CAPES, FAPERJ, and FAPESP (Brazil); MES (Bulgaria); CERN; CAS, MoST, and NSFC (China); COLCIENCIAS (Colombia); MSES and CSF (Croatia); RPF (Cyprus); MoER, ERC IUT and ERDF (Estonia); Academy of Finland, MEC, and HIP (Finland); CEA and CNRS/IN2P3 (France); BMBF, DFG, and HGF (Germany); GSRT (Greece); OTKA and NIH (Hungary); DAE and DST (India); IPM (Iran); SFI (Ireland); INFN (Italy); MSIP and NRF (Republic of Korea); LAS (Lithuania); MOE and UM (Malaysia); BUAP, CINVESTAV, CONACYT, LNS, SEP, and UASLP-FAI (Mexico); MBIE (New Zealand); PAEC (Pakistan); MSHE and NSC (Poland); FCT (Portugal); JINR (Dubna); MON, RosAtom, RAS and RFBR (Russia); MESTD (Serbia); SEIDI and CPAN (Spain); Swiss Funding Agencies (Switzerland); MST (Taipei); ThEPCenter, IPST, STAR and NSTDA (Thailand); TUBITAK and TAEK (Turkey); NASU and SFFR (Ukraine); STFC (United Kingdom); DOE and NSF (USA).

Individuals have received support from the Marie-Curie program and the European Research Council and EPLANET (European Union); the Leventis Foundation; the A. P. Sloan Foundation; the Alexander von Humboldt Foundation; the Belgian Federal Science Policy Office; the Fonds pour la Formation \`a la Recherche dans l'Industrie et dans l'Agriculture (FRIA-Belgium); the Agentschap voor Innovatie door Wetenschap en Technologie (IWT-Belgium); the Ministry of Education, Youth and Sports (MEYS) of the Czech Republic; the Council of Science and Industrial Research, India; the HOMING PLUS program of the Foundation for Polish Science, cofinanced from European Union, Regional Development Fund; the Mobility Plus program of the Ministry of Science and Higher Education (Poland); the OPUS program of the National Science Center (Poland); the Thalis and Aristeia programs cofinanced by EU-ESF and the Greek NSRF; the National Priorities Research Program by Qatar National Research Fund; the Programa Clar\'in-COFUND del Principado de Asturias; the Rachadapisek Sompot Fund for Postdoctoral Fellowship, Chulalongkorn University (Thailand); the Chulalongkorn Academic into Its 2nd Century Project Advancement Project (Thailand); and the Welch Foundation, contract C-1845.
\end{acknowledgments}
\vspace*{5em}
\bibliography{auto_generated}
\cleardoublepage \appendix\section{The CMS Collaboration \label{app:collab}}\begin{sloppypar}\hyphenpenalty=5000\widowpenalty=500\clubpenalty=5000\input{HIN-15-006-authorlist.tex}\end{sloppypar}
\end{document}

%% file: HIN-15-006-authorlist.tex
\textbf{Yerevan Physics Institute,  Yerevan,  Armenia}\\*[0pt]
V.~Khachatryan, A.M.~Sirunyan, A.~Tumasyan
\vskip\cmsinstskip
\textbf{Institut f\"{u}r Hochenergiephysik der OeAW,  Wien,  Austria}\\*[0pt]
W.~Adam, E.~Asilar, T.~Bergauer, J.~Brandstetter, E.~Brondolin, M.~Dragicevic, J.~Er\"{o}, M.~Flechl, M.~Friedl, R.~Fr\"{u}hwirth\cmsAuthorMark{1}, V.M.~Ghete, C.~Hartl, N.~H\"{o}rmann, J.~Hrubec, M.~Jeitler\cmsAuthorMark{1}, A.~K\"{o}nig, M.~Krammer\cmsAuthorMark{1}, I.~Kr\"{a}tschmer, D.~Liko, T.~Matsushita, I.~Mikulec, D.~Rabady, N.~Rad, B.~Rahbaran, H.~Rohringer, J.~Schieck\cmsAuthorMark{1}, J.~Strauss, W.~Treberer-Treberspurg, W.~Waltenberger, C.-E.~Wulz\cmsAuthorMark{1}
% ...
\vskip\cmsinstskip
% ...
1:~~Also at Vienna University of Technology, Vienna, Austria\\
% ...


%% file: HIN-15-006_temp.bbl
\providecommand{\href}[2]{#2}\begingroup\raggedright\begin{thebibliography}{10}%
\makeatletter
\providecommand{\hrefCMSnoop }[0]{\@secondoftwo}%
\makeatother
\providecommand{\doi}{\texttt{doi:}\begingroup \urlstyle{tt}\Url}

\bibitem{Andersen:1998vu}
\hrefCMSnoop {}{E.~Andersen {et~al.}, ``{Enhancement of central $\Lambda$,
  $\Xi$ and $\Omega$ yields in Pb-Pb collisions at 158 A GeV/$c$}'',} \textit{
  Phys. Lett. B} \textbf{ 433} (1998) 209,
\href{http://dx.doi.org/10.1016/S0370-2693(98)00689-3}{\doi{10.1016/S0370-2693(98)00689-3}}.

\bibitem{STAR}
\hrefCMSnoop {}{{STAR} Collaboration, ``{Experimental and theoretical
  challenges in the search for the quark gluon plasma: The STAR Collaboration's
  critical assessment of the evidence from RHIC collisions}'',} \textit{ Nucl.
  Phys. A} \textbf{ 757} (2005) 102,
  \href{http://dx.doi.org/10.1016/j.nuclphysa.2005.03.085}{\doi{10.1016/j.nuclphysa.2005.03.085}},
\href{http://www.arXiv.org/abs/nucl-ex/0501009}{\texttt{arXiv:nucl-ex/0501009}}.

\bibitem{Rafelski:1982pu}
\hrefCMSnoop {}{J.~Rafelski and B.~Muller, ``Strangeness Production in the
  Quark-Gluon Plasma'',} \textit{ Phys. Rev. Lett.} \textbf{ 48} (1982) 1066,
\href{http://dx.doi.org/10.1103/PhysRevLett.48.1066}{\doi{10.1103/PhysRevLett.48.1066}}.

\bibitem{Abelev:2007xp}
\hrefCMSnoop {}{{STAR} Collaboration, ``{Enhanced strange baryon production in
  Au + Au collisions compared to p + p at s(NN)**(1/2) = 200-GeV}'',} \textit{
  Phys. Rev. C} \textbf{ 77} (2008) 044908,
  \href{http://dx.doi.org/10.1103/PhysRevC.77.044908}{\doi{10.1103/PhysRevC.77.044908}},
\href{http://www.arXiv.org/abs/0705.2511}{\texttt{arXiv:0705.2511}}.

\bibitem{Andersen:1999ym}
\hrefCMSnoop {}{{WA97} Collaboration, ``{Strangeness enhancement at
  mid-rapidity in Pb Pb collisions at 158-A-GeV/c}'',} \textit{ Phys. Lett. B}
  \textbf{ 449} (1999) 401,
\href{http://dx.doi.org/10.1016/S0370-2693(99)00140-9}{\doi{10.1016/S0370-2693(99)00140-9}}.

\bibitem{Adams:2003fy}
\hrefCMSnoop {}{{STAR} Collaboration, ``{Multistrange baryon production in
  Au-Au collisions at S(NN)**1/2 = 130 GeV}'',} \textit{ Phys. Rev. Lett.}
  \textbf{ 92} (2004) 182301,
  \href{http://dx.doi.org/10.1103/PhysRevLett.92.182301}{\doi{10.1103/PhysRevLett.92.182301}},
\href{http://www.arXiv.org/abs/nucl-ex/0307024}{\texttt{arXiv:nucl-ex/0307024}}.

\bibitem{PHENIX}
\hrefCMSnoop {}{{PHENIX} Collaboration, ``{Formation of dense partonic matter
  in relativistic nucleus-nucleus collisions at RHIC: Experimental evaluation
  by the PHENIX Collaboration}'',} \textit{ Nucl. Phys. A} \textbf{ 757} (2005)
  184,
  \href{http://dx.doi.org/10.1016/j.nuclphysa.2005.03.086}{\doi{10.1016/j.nuclphysa.2005.03.086}},
\href{http://www.arXiv.org/abs/nucl-ex/0410003}{\texttt{arXiv:nucl-ex/0410003}}.

\bibitem{Khachatryan:2010gv}
\hrefCMSnoop {}{{CMS Collaboration}, ``{Observation of long-range near-side
  angular correlations in proton-proton collisions at the LHC}'',} \textit{
  JHEP} \textbf{ 09} (2010) 091,
  \href{http://dx.doi.org/10.1007/JHEP09(2010)091}{\doi{10.1007/JHEP09(2010)091}},
\href{http://www.arXiv.org/abs/1009.4122}{\texttt{arXiv:1009.4122}}.

\bibitem{Khachatryan:2012dih}
\hrefCMSnoop {}{{CMS Collaboration}, ``{Observation of long-range, near-side
  angular correlations in pPb collisions at the LHC}'',} \textit{ Phys. Lett.
  B} \textbf{ 718} (2013) 795,
  \href{http://dx.doi.org/10.1016/j.physletb.2012.11.025}{\doi{10.1016/j.physletb.2012.11.025}},
\href{http://www.arXiv.org/abs/1210.5482}{\texttt{arXiv:1210.5482}}.

\bibitem{atlas:2012fa}
\hrefCMSnoop {}{{ATLAS Collaboration}, ``{Observation of associated near-side
  and away-side long-range correlations in $\sqrt{s_{\mathrm{NN}}} = 5.02$\TeV
  proton-lead collisions with the ATLAS detector}'',} \textit{ Phys. Rev.
  Lett.} \textbf{ 110} (2013) 182302,
  \href{http://dx.doi.org/10.1103/PhysRevLett.110.182302}{\doi{10.1103/PhysRevLett.110.182302}},
\href{http://www.arXiv.org/abs/1212.5198}{\texttt{arXiv:1212.5198}}.

\bibitem{alice:2012qe}
\hrefCMSnoop {}{{ALICE Collaboration}, ``{Long-range angular correlations on
  the near and away side in \pPb\ collisions at $\sqrt{s_{\mathrm{NN}}} =
  5.02$\TeV}'',} \textit{ Phys. Lett. B} \textbf{ 719} (2013) 29,
  \href{http://dx.doi.org/10.1016/j.physletb.2013.01.012}{\doi{10.1016/j.physletb.2013.01.012}},
\href{http://www.arXiv.org/abs/1212.2001}{\texttt{arXiv:1212.2001}}.

\bibitem{Dusling:2015gta}
\hrefCMSnoop {}{K.~Dusling, W.~Li, and B.~Schenke, ``Novel collective phenomena
  in high-energy proton-proton and proton-nucleus collisions'',} \textit{ Int.
  J. Mod. Phys. E} \textbf{ 25} (2016), no.~01, 1630002,
  \href{http://dx.doi.org/10.1142/S0218301316300022}{\doi{10.1142/S0218301316300022}},
\href{http://www.arXiv.org/abs/1509.07939}{\texttt{arXiv:1509.07939}}.

\bibitem{Bozek:2011if}
\hrefCMSnoop {}{P.~Bo$\dot{\text{z}}$ek, ``{Collective flow in $p$-Pb and
  $d$-Pb collisions at TeV energies}'',} \textit{ Phys. Rev. C} \textbf{ 85}
  (2012) 014911,
  \href{http://dx.doi.org/10.1103/PhysRevC.85.014911}{\doi{10.1103/PhysRevC.85.014911}},
\href{http://www.arXiv.org/abs/1112.0915}{\texttt{arXiv:1112.0915}}.

\bibitem{Bozek:2012gr}
\hrefCMSnoop {}{P.~Bo$\dot{\text{z}}$ek and W.~Broniowski, ``{Correlations from
  hydrodynamic flow in \pPb\ collisions}'',} \textit{ Phys. Lett. B} \textbf{
  718} (2013) 1557,
  \href{http://dx.doi.org/10.1016/j.physletb.2012.12.051}{\doi{10.1016/j.physletb.2012.12.051}},
\href{http://www.arXiv.org/abs/1211.0845}{\texttt{arXiv:1211.0845}}.

\bibitem{Bzdak:2013zma}
\hrefCMSnoop {}{A.~Bzdak, B.~Schenke, P.~Tribedy, and R.~Venugopalan,
  ``{Initial state geometry and the role of hydrodynamics in proton-proton,
  proton-nucleus and deuteron-nucleus collisions}'',} \textit{ Phys. Rev. C}
  \textbf{ 87} (2013) 064906,
  \href{http://dx.doi.org/10.1103/PhysRevC.87.064906}{\doi{10.1103/PhysRevC.87.064906}},
\href{http://www.arXiv.org/abs/1304.3403}{\texttt{arXiv:1304.3403}}.

\bibitem{Werner:2010ss}
\hrefCMSnoop {}{K.~Werner, I.~Karpenko, and T.~Pierog, ````Ridge'' in
  Proton-Proton Scattering at {7 TeV}'',} \textit{ Phys. Rev. Lett.} \textbf{
  106} (2011) 122004,
  \href{http://dx.doi.org/10.1103/PhysRevLett.106.122004}{\doi{10.1103/PhysRevLett.106.122004}},
\href{http://www.arXiv.org/abs/1011.0375}{\texttt{arXiv:1011.0375}}.

\bibitem{Dusling:2012wy}
\hrefCMSnoop {}{K.~Dusling and R.~Venugopalan, ``{Explanation of systematics of
  CMS p+Pb high multiplicity di-hadron data at $\sqrt{s_{\mathrm{NN}}} = 5.02$
  TeV}'',} \textit{ Phys. Rev. D} \textbf{ 87} (2013) 054014,
  \href{http://dx.doi.org/10.1103/PhysRevD.87.054014}{\doi{10.1103/PhysRevD.87.054014}},
\href{http://www.arXiv.org/abs/1211.3701}{\texttt{arXiv:1211.3701}}.

\bibitem{Dusling:2012cg}
\hrefCMSnoop {}{K.~Dusling and R.~Venugopalan, ``{Evidence for BFKL and
  saturation dynamics from dihadron spectra at the LHC}'',} \textit{ Phys. Rev.
  D} \textbf{ 87} (2013) 051502,
  \href{http://dx.doi.org/10.1103/PhysRevD.87.051502}{\doi{10.1103/PhysRevD.87.051502}},
\href{http://www.arXiv.org/abs/1210.3890}{\texttt{arXiv:1210.3890}}.

\bibitem{Dumitru:2014yza}
\hrefCMSnoop {}{A.~Dumitru, L.~McLerran, and V.~Skokov, ``Azimuthal asymmetries
  and the emergence of ``collectivity'' from multi-particle correlations in
  high-energy {pA} collisions'',} \textit{ Phys. Lett. B} \textbf{ 743} (2015)
  134,
  \href{http://dx.doi.org/10.1016/j.physletb.2015.02.046}{\doi{10.1016/j.physletb.2015.02.046}},
\href{http://www.arXiv.org/abs/1410.4844}{\texttt{arXiv:1410.4844}}.

\bibitem{Gyulassy:2014cfa}
\hrefCMSnoop {}{M.~Gyulassy, P.~Levai, I.~Vitev, and T.~S. Biro, ``{Non-Abelian
  bremsstrahlung and azimuthal asymmetries in high energy $p+A$ reactions}'',}
  \textit{ Phys. Rev. D} \textbf{ 90} (2014) 054025,
  \href{http://dx.doi.org/10.1103/PhysRevD.90.054025}{\doi{10.1103/PhysRevD.90.054025}},
\href{http://www.arXiv.org/abs/1405.7825}{\texttt{arXiv:1405.7825}}.

\bibitem{Li:2012hc}
\hrefCMSnoop {}{W.~Li, ``{Observation of a 'ridge' correlation structure in
  high multiplicity proton-proton collisions: A brief review}'',} \textit{ Mod.
  Phys. Lett. A} \textbf{ 27} (2012) 1230018,
  \href{http://dx.doi.org/10.1142/S0217732312300182}{\doi{10.1142/S0217732312300182}},
\href{http://www.arXiv.org/abs/1206.0148}{\texttt{arXiv:1206.0148}}.

\bibitem{He:2015hfa}
L.~He\hrefCMSnoop {}{ {et~al.}, ``{Anisotropic parton escape is the dominant
  source of azimuthal anisotropy in transport models}'',} \textit{ Phys. Lett.
  B} \textbf{ 753} (2016) 506,
  \href{http://dx.doi.org/10.1016/j.physletb.2015.12.051}{\doi{10.1016/j.physletb.2015.12.051}},
\href{http://www.arXiv.org/abs/1502.05572}{\texttt{arXiv:1502.05572}}.

\bibitem{Abelev:2013vea}
\hrefCMSnoop {}{{ALICE Collaboration}, ``{Centrality dependence of $\pi$, $K$,
  $p$ production in Pb-Pb collisions at $\sqrt{s_{\mathrm{NN}}}$ = 2.76
  TeV}'',} \textit{ Phys. Rev. C} \textbf{ 88} (2013) 044910,
  \href{http://dx.doi.org/10.1103/PhysRevC.88.044910}{\doi{10.1103/PhysRevC.88.044910}},
\href{http://www.arXiv.org/abs/1303.0737}{\texttt{arXiv:1303.0737}}.

\bibitem{Abelev:2014pua}
\hrefCMSnoop {}{{ALICE Collaboration}, ``{Elliptic flow of identified hadrons
  in Pb-Pb collisions at $ \sqrt{s_{\mathrm{NN}}}=2.76 $ TeV}'',} \textit{
  JHEP} \textbf{ 06} (2015) 190,
  \href{http://dx.doi.org/10.1007/JHEP06(2015)190}{\doi{10.1007/JHEP06(2015)190}},
\href{http://www.arXiv.org/abs/1405.4632}{\texttt{arXiv:1405.4632}}.

\bibitem{Song:2013qma}
\hrefCMSnoop {}{H.~Song, S.~Bass, and U.~W. Heinz, ``{Spectra and elliptic flow
  for identified hadrons in 2.76A TeV Pb+Pb collisions}'',} \textit{ Phys. Rev.
  C} \textbf{ 89} (2014) 034919,
  \href{http://dx.doi.org/10.1103/PhysRevC.89.034919}{\doi{10.1103/PhysRevC.89.034919}},
\href{http://www.arXiv.org/abs/1311.0157}{\texttt{arXiv:1311.0157}}.

\bibitem{Zhu:2015dfa}
\hrefCMSnoop {}{X.~Zhu, F.~Meng, H.~Song, and Y.-X. Liu, ``{Hybrid model
  approach for strange and multistrange hadrons in 2.76A TeV Pb+Pb
  collisions}'',} \textit{ Phys. Rev. C} \textbf{ 91} (2015), no.~3, 034904,
  \href{http://dx.doi.org/10.1103/PhysRevC.91.034904}{\doi{10.1103/PhysRevC.91.034904}},
\href{http://www.arXiv.org/abs/1501.03286}{\texttt{arXiv:1501.03286}}.

\bibitem{Abelev:2013haa}
\hrefCMSnoop {}{{ALICE Collaboration}, ``Multiplicity dependence of pion, kaon,
  proton and lambda production in p-Pb collisions at {$\sqrt{s_{\mathrm{NN}}}$
  = 5.02 TeV}'',} \textit{ Phys. Lett. B} \textbf{ 728} (2014) 25,
  \href{http://dx.doi.org/10.1016/j.physletb.2013.11.020}{\doi{10.1016/j.physletb.2013.11.020}},
\href{http://www.arXiv.org/abs/1307.6796}{\texttt{arXiv:1307.6796}}.

\bibitem{Chatrchyan:2013eya}
\hrefCMSnoop {}{{CMS Collaboration}, ``Study of the production of charged
  pions, kaons, and protons in {pPb} collisions at {$\sqrt{s_{\mathrm{NN}}} =
  5.02$ {TeV}}'',} \textit{ Eur. Phys. J. C} \textbf{ 74} (2014) 2847,
  \href{http://dx.doi.org/10.1140/epjc/s10052-014-2847-x}{\doi{10.1140/epjc/s10052-014-2847-x}},
\href{http://www.arXiv.org/abs/1307.3442}{\texttt{arXiv:1307.3442}}.

\bibitem{ABELEV:2013wsa}
\hrefCMSnoop {}{{ALICE Collaboration}, ``{Long-range angular correlations of
  $\pi$, K and p in p--Pb collisions at $\sqrt{s_{\mathrm{NN}}} =
  5.02$\TeV}'',} \textit{ Phys. Lett. B} \textbf{ 726} (2013) 164,
  \href{http://dx.doi.org/10.1016/j.physletb.2013.08.024}{\doi{10.1016/j.physletb.2013.08.024}},
\href{http://www.arXiv.org/abs/1307.3237}{\texttt{arXiv:1307.3237}}.

\bibitem{Khachatryan:2014jra}
\hrefCMSnoop {}{{CMS Collaboration}, ``{Long-range two-particle correlations of
  strange hadrons with charged particles in pPb and PbPb collisions at LHC
  energies}'',} \textit{ Phys. Lett. B} \textbf{ 742} (2015) 200,
  \href{http://dx.doi.org/10.1016/j.physletb.2015.01.034}{\doi{10.1016/j.physletb.2015.01.034}},
\href{http://www.arXiv.org/abs/1409.3392}{\texttt{arXiv:1409.3392}}.

\bibitem{Shuryak:2013ke}
\hrefCMSnoop {}{E.~Shuryak and I.~Zahed, ``{High-multiplicity $pp$ and $pA$
  collisions: Hydrodynamics at its edge}'',} \textit{ Phys. Rev. C} \textbf{
  88} (2013) 044915,
  \href{http://dx.doi.org/10.1103/PhysRevC.88.044915}{\doi{10.1103/PhysRevC.88.044915}},
\href{http://www.arXiv.org/abs/1301.4470}{\texttt{arXiv:1301.4470}}.

\bibitem{Schnedermann_blastwave}
\hrefCMSnoop {}{E.~Schnedermann, J.~Sollfrank, and U.~Heinz, ``Thermal
  phenomenology of hadrons from {200 $A$ GeV S+S} collisions'',} \textit{ Phys.
  Rev. C} \textbf{ 48} (1993) 2462,
  \href{http://dx.doi.org/10.1103/PhysRevC.48.2462}{\doi{10.1103/PhysRevC.48.2462}},
  \href{http://www.arXiv.org/abs/nucl-th/9307020}{\texttt{arXiv:nucl-th/9307020}}.

\bibitem{Albajar:1989an}
\hrefCMSnoop {}{{UA1} Collaboration, ``A study of the general characteristics
  of proton-antiproton collisions at {$\sqrt{s} = 0.2$ TeV to 0.9 TeV}'',}
  \textit{ Nucl. Phys. B} \textbf{ 335} (1990) 261,
\href{http://dx.doi.org/10.1016/0550-3213(90)90493-W}{\doi{10.1016/0550-3213(90)90493-W}}.

\bibitem{Alexopoulos:1988na}
\hrefCMSnoop {}{T.~Alexopoulos {et~al.}, ``Multiplicity dependence of the
  transverse momentum spectrum for centrally produced hadrons in
  antiproton-proton collisions at {$\sqrt{s}=1.8$ Tev}'',} \textit{ Phys. Rev.
  Lett.} \textbf{ 60} (1988) 1622,
\href{http://dx.doi.org/10.1103/PhysRevLett.60.1622}{\doi{10.1103/PhysRevLett.60.1622}}.

\bibitem{Abelev:2008ab}
\hrefCMSnoop {}{{STAR} Collaboration, ``Systematic measurements of identified
  particle spectra in $p p$, $d+\mathrm{Au}$ and $\mathrm{Au}+\mathrm{Au}$
  collisions from {STAR}'',} \textit{ Phys. Rev. C} \textbf{ 79} (2009) 034909,
  \href{http://dx.doi.org/10.1103/PhysRevC.79.034909}{\doi{10.1103/PhysRevC.79.034909}},
\href{http://www.arXiv.org/abs/0808.2041}{\texttt{arXiv:0808.2041}}.

\bibitem{Ortiz:2013yxa}
A.~Ortiz~Velasquez\hrefCMSnoop {}{ {et~al.}, ``Color Reconnection and Flowlike
  Patterns in $pp$ Collisions'',} \textit{ Phys. Rev. Lett.} \textbf{ 111}
  (2013) 042001,
  \href{http://dx.doi.org/10.1103/PhysRevLett.111.042001}{\doi{10.1103/PhysRevLett.111.042001}},
\href{http://www.arXiv.org/abs/1303.6326}{\texttt{arXiv:1303.6326}}.

\bibitem{Bozek:2013sda}
\hrefCMSnoop {}{P.~Bo$\dot{\text{z}}$ek, A.~Bzdak, and V.~Skokov, ``{The
  rapidity dependence of the average transverse momentum in p+Pb collisions at
  the LHC: the color glass condensate versus hydrodynamics}'',} \textit{ Phys.
  Lett. B} \textbf{ 728} (2014) 662,
  \href{http://dx.doi.org/10.1016/j.physletb.2013.12.034}{\doi{10.1016/j.physletb.2013.12.034}},
\href{http://www.arXiv.org/abs/1309.7358}{\texttt{arXiv:1309.7358}}.

\bibitem{TRK-11-001}
\hrefCMSnoop {}{{CMS Collaboration}, ``{Description and performance of track
  and primary-vertex reconstruction with the CMS tracker}'',} \textit{ JINST}
  \textbf{ 9} (2014) P10009,
  \href{http://dx.doi.org/10.1088/1748-0221/9/10/P10009}{\doi{10.1088/1748-0221/9/10/P10009}},
\href{http://www.arXiv.org/abs/1405.6569}{\texttt{arXiv:1405.6569}}.

\bibitem{Chatrchyan:2008zzk}
\hrefCMSnoop {}{{CMS Collaboration}, ``The {CMS} experiment at the {CERN}
  {LHC}'',} \textit{ JINST} \textbf{ 3} (2008) S08004,
\href{http://dx.doi.org/10.1088/1748-0221/3/08/S08004}{\doi{10.1088/1748-0221/3/08/S08004}}.

\bibitem{GEANT4}
\hrefCMSnoop {}{{Geant4} Collaboration, ``{Geant4: A simulation toolkit}'',}
  \textit{ Nucl. Instrum. and Methods A} \textbf{ 506} (2003) 250,
\href{http://dx.doi.org/10.1016/S0168-9002(03)01368-8}{\doi{10.1016/S0168-9002(03)01368-8}}.

\bibitem{Chatrchyan:2013nka}
\hrefCMSnoop {}{{CMS} Collaboration, ``{Multiplicity and transverse momentum
  dependence of two- and four-particle correlations in \pPb\ and \PbPb\
  collisions}'',} \textit{ Phys. Lett. B} \textbf{ 724} (2013) 213,
  \href{http://dx.doi.org/10.1016/j.physletb.2013.06.028}{\doi{10.1016/j.physletb.2013.06.028}},
\href{http://www.arXiv.org/abs/1305.0609}{\texttt{arXiv:1305.0609}}.

\bibitem{Chatrchyan:2012xq}
\hrefCMSnoop {}{{CMS Collaboration}, ``{Azimuthal anisotropy of charged
  particles at high transverse momenta in PbPb collisions at $\rootsNN =
  2.76\TeV$}'',} \textit{ Phys. Rev. Lett.} \textbf{ 109} (2012) 022301,
  \href{http://dx.doi.org/10.1103/PhysRevLett.109.022301}{\doi{10.1103/PhysRevLett.109.022301}},
\href{http://www.arXiv.org/abs/1204.1850}{\texttt{arXiv:1204.1850}}.

\bibitem{Adam:2005zf}
\hrefCMSnoop {}{{CMS Trigger and Data Acquisition Group} Collaboration, ``{The
  CMS high level trigger}'',} \textit{ Eur. Phys. J. C} \textbf{ 46} (2006)
  605,
  \href{http://dx.doi.org/10.1140/epjc/s2006-02495-8}{\doi{10.1140/epjc/s2006-02495-8}},
\href{http://www.arXiv.org/abs/hep-ex/0512077}{\texttt{arXiv:hep-ex/0512077}}.

\bibitem{Pierog:2013ria}
T.~Pierog\hrefCMSnoop {}{ {et~al.}, ``{EPOS LHC}: Test of collective
  hadronization with data measured at the {CERN Large Hadron Collider}'',}
  \textit{ Phys. Rev. C} \textbf{ 92} (2015) 034906,
  \href{http://dx.doi.org/10.1103/PhysRevC.92.034906}{\doi{10.1103/PhysRevC.92.034906}}.

\bibitem{Gyulassy:1994ew}
\hrefCMSnoop {}{M.~Gyulassy and X.-N. Wang, ``{HIJING 1.0: A Monte Carlo
  program for parton and particle production in high-energy hadronic and
  nuclear collisions}'',} \textit{ Comput. Phys. Commun.} \textbf{ 83} (1994)
  307,
  \href{http://dx.doi.org/10.1016/0010-4655(94)90057-4}{\doi{10.1016/0010-4655(94)90057-4}},
\href{http://www.arXiv.org/abs/nucl-th/9502021}{\texttt{arXiv:nucl-th/9502021}}.

\bibitem{Sjostrand:2006za}
\hrefCMSnoop {}{T.~Sj{\"o}strand, S.~Mrenna, and P.~Skands, ``{PYTHIA} 6.4
  physics and manual'',} \textit{ JHEP} \textbf{ 05} (2006) 026,
  \href{http://dx.doi.org/10.1088/1126-6708/2006/05/026}{\doi{10.1088/1126-6708/2006/05/026}},
\href{http://www.arXiv.org/abs/hep-ph/0603175}{\texttt{arXiv:hep-ph/0603175}}.

\bibitem{Sjostrand:2007gs}
\hrefCMSnoop {}{T.~Sj{\"o}strand, S.~Mrenna, and P.~Skands, ``A brief
  introduction to {PYTHIA} 8.1'',} \textit{ Comp. Phys. Comm.} \textbf{ 178}
  (2008) 852,
  \href{http://dx.doi.org/10.1016/j.cpc.2008.01.036}{\doi{10.1016/j.cpc.2008.01.036}},
\href{http://www.arXiv.org/abs/0710.3820}{\texttt{arXiv:0710.3820}}.

\bibitem{CMS-PAS-TRK-10-002}
\href {http://cdsweb.cern.ch/record/1279139}{{CMS Collaboration}, ``Measurement
  of Tracking Efficiency'',} CMS Physics Analysis Summary CMS-PAS-TRK-10-002,
  2010.

\bibitem{Khachatryan:2011tm}
\hrefCMSnoop {}{{CMS Collaboration}, ``Strange particle production in pp
  collisions at $\roots = 0.9$ and {7\TeV}'',} \textit{ JHEP} \textbf{ 05}
  (2011) 064,
  \href{http://dx.doi.org/10.1007/JHEP05(2011)064}{\doi{10.1007/JHEP05(2011)064}},
\href{http://www.arXiv.org/abs/1102.4282}{\texttt{arXiv:1102.4282}}.

\bibitem{PDG}
\hrefCMSnoop {}{{Particle Data Group}, K.~A. Olive {et~al.}, ``{Review of
  Particle Physics}'',} \textit{ Chin. Phys. C} \textbf{ 38} (2014) 090001,
\href{http://dx.doi.org/10.1088/1674-1137/38/9/090001}{\doi{10.1088/1674-1137/38/9/090001}}.

\bibitem{Bozek:2012qs}
\hrefCMSnoop {}{P.~Bo$\dot{\text{z}}$ek and I.~Wyskiel-Piekarska, ``{Particle
  spectra in Pb-Pb collisions at $\sqrt{s_{\mathrm{NN}}} =\ $ 2.76 $\,\text
  {TeV}$}'',} \textit{ Phys. Rev. C} \textbf{ 85} (2012) 064915,
  \href{http://dx.doi.org/10.1103/PhysRevC.85.064915}{\doi{10.1103/PhysRevC.85.064915}},
\href{http://www.arXiv.org/abs/1203.6513}{\texttt{arXiv:1203.6513}}.

\bibitem{Karpenko:2012yf}
\hrefCMSnoop {}{I.~A. Karpenko, {\relax Yu}.~M. Sinyukov, and K.~Werner,
  ``{Uniform description of bulk observables in the hydrokinetic model of $A+A$
  collisions at the BNL Relativistic Heavy Ion Collider and the CERN Large
  Hadron Collider}'',} \textit{ Phys. Rev. C} \textbf{ 87} (2013) 024914,
  \href{http://dx.doi.org/10.1103/PhysRevC.87.024914}{\doi{10.1103/PhysRevC.87.024914}},
\href{http://www.arXiv.org/abs/1204.5351}{\texttt{arXiv:1204.5351}}.

\bibitem{Fries:2008hs}
\hrefCMSnoop {}{R.~J. Fries, V.~Greco, and P.~Sorensen, ``{Coalescence Models
  For Hadron Formation From Quark Gluon Plasma}'',} \textit{ Ann. Rev. Nucl.
  Part. Sci.} \textbf{ 58} (2008) 177,
  \href{http://dx.doi.org/10.1146/annurev.nucl.58.110707.171134}{\doi{10.1146/annurev.nucl.58.110707.171134}},
\href{http://www.arXiv.org/abs/0807.4939}{\texttt{arXiv:0807.4939}}.

\bibitem{Chatrchyan:2012qb}
\hrefCMSnoop {}{{CMS Collaboration}, ``Study of the inclusive production of
  charged pions, kaons, and protons in pp collisions at $\sqrt{s}=0.9$, 2.76,
  and 7 {TeV}'',} \textit{ Eur. Phys. J. C} \textbf{ 72} (2012) 2164,
  \href{http://dx.doi.org/10.1140/epjc/s10052-012-2164-1}{\doi{10.1140/epjc/s10052-012-2164-1}},
\href{http://www.arXiv.org/abs/1207.4724}{\texttt{arXiv:1207.4724}}.

\bibitem{Adam:2016bpr}
\hrefCMSnoop {}{{ALICE Collaboration}, ``{Production of K$^{*}$ (892)$^{0}$ and
  $\phi $ (1020) in p-Pb collisions at $\sqrt{s_{{\text {NN}}}}$ = 5.02
  TeV}'',} \textit{ Eur. Phys. J. C} \textbf{ 76} (2016), no.~5, 245,
  \href{http://dx.doi.org/10.1140/epjc/s10052-016-4088-7}{\doi{10.1140/epjc/s10052-016-4088-7}},
\href{http://www.arXiv.org/abs/1601.07868}{\texttt{arXiv:1601.07868}}.

\end{thebibliography}\endgroup
